\documentclass[english,aps,reprint]{revtex4-2}
\usepackage[latin9]{inputenc}
\usepackage{color}
\usepackage{babel}
\usepackage{mathtools}
\usepackage{amsmath}
\usepackage{amsthm}
\usepackage{amssymb}
\usepackage{graphicx}
\usepackage[unicode=true,
 bookmarks=false,
 breaklinks=false,pdfborder={0 0 1},colorlinks=true]
 {hyperref}
\hypersetup{
 linkcolor=black,filecolor=magenta,urlcolor=black,citecolor=black}

\makeatletter

\usepackage[usenames,dvipsnames,svgnames,table]{xcolor}

\usepackage{chemformula}

\usepackage{amsfonts}
\usepackage{amsthm}\usepackage{bm}

\usepackage{mathrsfs}
\usepackage{physics}
\usepackage{dsfont}

\DeclareMathOperator\arctanh{arctanh}

\usepackage{amsfonts}

\usepackage{standalone}

\makeatother

\begin{document}
\title{The role of electromagnetic gauge-field fluctuations in the selection
between chiral and nematic superconductivity}
\author{Virginia Gali}
\author{Rafael M. Fernandes}
\affiliation{School of Physics and Astronomy, University of Minnesota, Minneapolis,
MN 55455, USA}
\date{\today}
\begin{abstract}
Motivated by the observation of nematic superconductivity in several
systems, we revisit the problem of the leading pairing instability
of two-component unconventional superconductors on the triangular
lattice -- such as $\left(p_{x},\,p_{y}\right)$-wave and $\left(d_{x^{2}-y^{2}},\,d_{xy}\right)$-wave.
Such a system has two possible superconducting states: the chiral
state (e.g. $p+ip$ or $d+id$), which breaks time-reversal symmetry,
and the nematic state (e.g. $p+p$ or $d+d$), which breaks the threefold
rotational symmetry of the lattice. Weak-coupling calculations generally
favor the chiral over the nematic superconducting state, raising the
question of what mechanism can stabilize the latter. Here, we show
that the electromagnetic field fluctuations can play a crucial role
in selecting between these two states. Specifically, we derive and
analyze the effective free energy for the two-component superconducting
order parameter after integrating out the gauge-field fluctuations,
which is formally justified if the spatial order parameter fluctuations
can be neglected. A non-analytic cubic term arises, as in the case
of a conventional $s$-wave superconductor. However, unlike the latter,
the cubic term depends on the relative phase and on the relative amplitudes
between the two order parameter components, in such a way that it
generally favors the nematic state. This result is a direct consequence
of the fact that the stiffness of the superconducting order parameter
is not isotropic. Competition with the quartic term, which favors
the chiral state, leads to a renormalized phase diagram in which the
nematic state displaces the chiral state over a wide region in the
parameter space. We analyze the stability of the fluctuation-induced
nematic phase, generalize our results to tetragonal lattices, and
discuss their applicability to candidate nematic superconductors,
including twisted bilayer graphene. 
\end{abstract}
\maketitle

\section{Introduction\label{sec:introduction}}

A nematic superconductor spontaneously breaks not only the $U(1)$
gauge symmetry, but also a discrete rotational symmetry of the system,
thus lowering the symmetry of the point group that characterizes the
underlying lattice. Recent experiments have reported evidence of rotational
symmetry-breaking superconducting states in different quantum materials,
such as doped Bi$_{2}$Se$_{3}$
\cite{MatanoK2016Ssbi,ShingoYonezawa2016Tefn,PanY.Rsbi,AsabaTomoya2017RSBi},
few-layer NbSe$_{2}$ \cite{hamill2020unexpected,cho2020distinct},
the topological semimetal CaSn$_{3}$ \cite{Siddiquee2022}, and the iron-based
superconductors Ba$_{1-x}$K$_{x}$Fe$_{2}$As$_{2}$ \cite{LiJun2017Nssi}
and LiFeAs \cite{Borisenko2020}. There is a longer list of materials in which
superconductivity can coexist with nematic order, such
as the iron chalcogenide FeSe \cite{Coldea15} or the nickel arsenide BaNi$_2$As$_2$ \cite{Eckberg2020}, but in these cases the superconducting
state emerges in the presence of a nematically ordered state that
onsets at much higher temperatures \cite{Fernandes2022}. Interestingly, the recently
discovered twisted bilayer graphene \cite{CaoYuan2018Usim,Cao2018,YankowitzMatthew2019Tsit,LuXiaobo2019Soma}
has also been reported to display a nematic superconducting state
in the ``hole-doped'' side of the phase diagram, as indicated by
the in-plane anisotropy of the critical magnetic field and of the
critical current \cite{cao2020nematicity}.

Theoretically, a nematic pairing state requires the simultaneous existence
of (at least) two superconducting order parameters whose relative
phase is not $\pi/2$. Generally, there are two different scenarios
in which this can happen. In the first case, two independent order
parameters, $\psi_{1}$ and $\psi_{2}$, condense at similar temperatures
due to some fine tuning of the microscopic parameters involved \cite{Willa2020}. One
example is the $s+d$ state proposed in Ba$_{1-x}$K$_{x}$Fe$_{2}$As$_{2}$ \cite{Fernandes_Millis, Livanas2015}. In the second scenario, the superconducting order
parameter has two symmetry-related components $\psi=\left(\psi_{1},\,\psi_{2}\right)$,
i.e. it transforms as a two-dimensional irreducible representation
(irrep) of the lattice point group. Examples of such order parameters
include the $\left(p_{x},\,p_{y}\right)$-wave and $\left(d_{x^{2}-y^{2}},\,d_{xy}\right)$-wave
in triangular lattices or the $\left(p_{x},\,p_{y}\right)$-wave
and the $\left(d_{xz},\,d_{yz}\right)$-wave in tetragonal lattices
\cite{SigristUeda1991}. Since this case does not require fine tuning, we will focus
on it in the remainder of the paper.

\begin{figure*}
\includegraphics{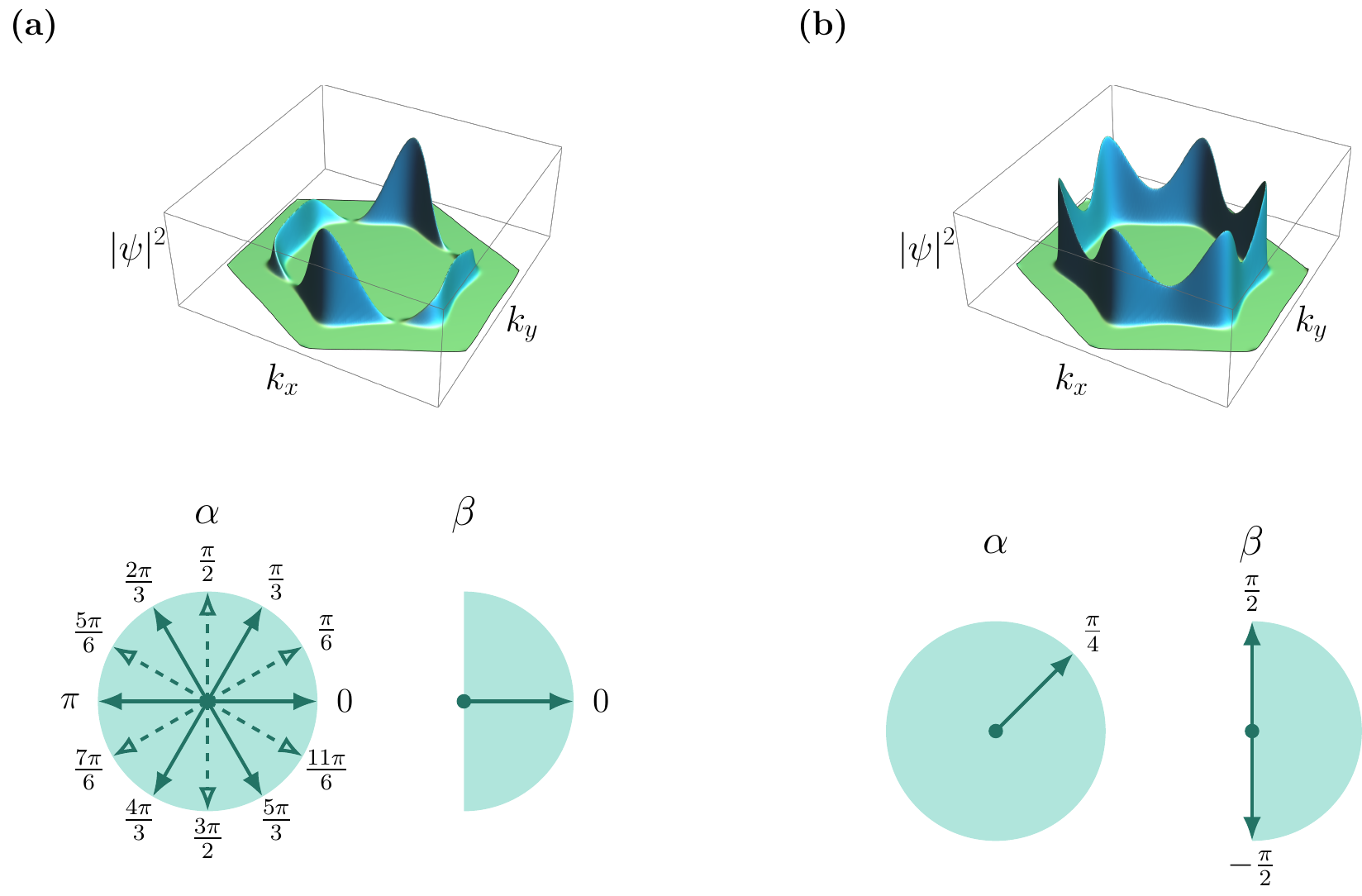} \vspace{-0.3cm}
 \caption{\label{fig:gaps}Plots of the superconducting gap function along a
generic Fermi surface on the triangular lattice. The order parameter
is parametrized in terms of three angles as $\psi=|\psi|e^{i\varphi}\left(\cos\alpha,e^{i\beta}\sin\alpha\right)$.
In this figure, the two components of $\psi$ transform as $d_{x^{2}-y^{2}}$
and $d_{xy}$ waves ($E_{2}$ irrep of $D_{6}$). \textbf{(a)} The
gap function in the nematic state, which lowers the $C_{6z}$ rotational
symmetry of the lattice to $C_{2z}$, is obtained when $\alpha=\tfrac{n\pi}{6}$
[with even $n=0,2,..,10$ (solid arrows) or odd $n=1,3,...,11$ (dashed arrows)] and $\beta=0$.
To produce this plot we chose $\alpha=0$. \textbf{(b)} The gap function
in the chiral state is obtained when $\alpha=\tfrac{\pi}{4}$ and
$\beta=\pm\tfrac{\pi}{2}$. It does not break any lattice symmetry.
However, this state breaks time reversal symmetry.}
\end{figure*}

It is convenient to parametrize $\psi$ in terms of three angles $\varphi\in[0,2\pi)$,
$\beta\in\left[-\pi/2,\,\pi/2\right]$ and $\alpha\in[0,2\pi)$ as
$\psi=|\psi|e^{i\varphi}\left(\cos{\alpha},e^{i\beta}\sin{\alpha}\right)$
\cite{Fernandes2019}. Below the transition temperature $T_{c}$, the global phase
$\varphi$ acquires a definite value and the $U(1)$ gauge symmetry
is broken. As for $\alpha$, which describes the relative amplitudes
between the two superconducting order parameters, and $\beta$, which
describes the relative phase between $\psi_{1}$ and $\psi_{2}$,
their allowed values are not continuous, but restricted to discrete
sets by the symmetries of the system. In the particular case of the
triangular (or honeycomb) lattice, there are two different possible
sets of values \cite{HeckerSchmalian2018}. The first one is $\beta=0$ and $\alpha=\tfrac{n\pi}{6}$
(with even $n=0,2,..,10$ or odd $n=1,3,...,11$), which corresponds
to a nematic superconducting state. Fig. \ref{fig:gaps}(a) shows
the absolute value square of the gap ($\left|\psi\right|^{2}$) in
the nematic phase, which clearly breaks the threefold rotational symmetry
$C_{3z}$ of the lattice. Note that there are points in which $\left|\psi\right|^{2}=0$,
corresponding to gap nodes. The different values of $\alpha$ correspond
to the different ways of breaking the $C_{3z}$ symmetry. The second
set of allowed values corresponds to $\alpha=\tfrac{\pi}{4}$ and
$\beta=\pm\tfrac{\pi}{2}$. Because $\psi^{*}\neq\psi$, time-reversal
symmetry is broken, and the superconducting state is chiral. In this
situation, $\left|\psi\right|^{2}$ respects the threefold rotational
symmetry of the lattice and is never zero, as shown in Fig. \ref{fig:gaps}(b).

The key question is which microscopic mechanisms are responsible for
the selection between the two possible pairing states -- nematic
or chiral. An argument usually invoked is that the chiral state should
be favored, since it completely gaps out the Fermi surface {[}Fig. \ref{fig:gaps}(b){]},
which would presumably maximize the condensation energy.
In agreement with this expectation, weak-coupling calculations find
that the chiral state is generally preferred \cite{KoziiVladyslav2019Nssb,BlackSchaffer2007,Nandkishore2012,Thomale2012,DHLee2012} -- unless spin-orbit coupling is significant \cite{Fu2014}. Moreover, in noncentrosymmetric systems, time-reversal symmetry must be broken \cite{Scheurer2017}. These results raise the interesting
question of which mechanism stabilizes the nematic superconducting
states that appear to be realized in the materials discussed above.
Besides the aforementioned possibility of nearly degenerate single-component
pairing states \citep{SZLin2018,ChichinadzeDmitryV2020Nsit,Scheurer2020,WangYuxuan2020Tans},
it has been pointed out that, in the case of a two-component superconductor,
coupling to strong normal-state nematic or density-wave fluctuations
can tip the balance in favor of nematic superconductivity \cite{KoziiVladyslav2019Nssb,Fernandes_Millis}.

In this paper, we discuss another possible mechanism that does not
require additional degrees of freedom or fine tuning. The key point
is that, because the superconducting order parameter is charged, it
couples to electromagnetic fluctuations. The effect of the gauge-field
fluctuations on conventional $s$-wave superconductors has been widely
investigated \citep{HalperinLubenskyMa1974,DasguptaHalperin1981,Kleinert1982,Herbut1996,Sudbo2002,Kleinert2003}. The seminal
work of Ref. \citep{HalperinLubenskyMa1974} showed that, upon integrating
out the gauge-field fluctuations, the superconducting transition becomes
weakly first-order due to the emergence of a non-analytic negative cubic term
in the free-energy expansion. Such an effect would be very small to
be detected due to the narrow window in which fluctuations are important
in $s$-wave superconductors. Because this procedure of integrating
out the electromagnetic fields is formally justified only when the
spatial order parameter fluctuations can be neglected, this conclusion
is robust for type-I superconductors. For type-II superconductors,
duality mappings and Monte Carlo simulations indicate that the transition
remains second-order \citep{DasguptaHalperin1981,Kleinert1982,Sudbo2002}.

The role of gauge-field fluctuations on layered unconventional superconductors,
where fluctuations generally can play a more prominent role than in
conventional superconductors, has been less studied. Ref.
\cite{Millev1990} considered the case of a general multi-band superconductor
with isotropic stiffness and found, like in the $s$-wave case, a
fluctuation-induced first-order transition via a renormalization-group
calculation. A similar result was found in Ref. \cite{LiBelitzToner2009} for a $p$-wave
superconductor, and Ref. \cite{Baym2004} reported the same outcome in the case of color superconductivity, where the gauge field is non-Abelian. Here, we extend this kind of perturbative analysis
to the case of two-component superconductors in triangular and tetragonal
lattices, taking into account the anisotropy of the superconducting
stiffness introduced by the crystal lattice. Specifically, we integrate
out the gauge-field fluctuations to obtain a renormalized Landau free-energy,
which is then minimized. 

Similarly to the $s$-wave \citep{HalperinLubenskyMa1974} and isotropic
unconventional superconductor cases \cite{Millev1990,LiBelitzToner2009},
we find a non-analytic cubic term with an overall negative sign, indicative
of a first-order transition. However, the main difference is that
this non-analytic term is not only dependent on $\left|\psi\right|^{3}$,
but also on the angles $\alpha$ and $\beta$ that distinguish between
the chiral and nematic states. This happens because the superconducting
stiffness is not isotropic, as in the $s$-wave case. Interestingly,
by combining numerical and analytical calculations, we find that the
cubic contribution to the free-energy is always minimized for the
nematic state. Consequently, because the chiral state arises from
the minimization of quartic terms of the free energy, the nematic
state becomes the global minimum of the renormalized free-energy in
a wide region of the parameter-space where the chiral state was the
global minimum of the mean-field free-energy. We further analyze the
stability of this gauge-field-fluctuations induced nematic state as
temperature is lowered below $T_{c}$. Finally, we discuss the limitations
of our approach and the possible application of our results to twisted bilayer graphene
and nematic superconductors in general.

The paper is organized as follows: we derive and solve the superconducting
free-energy renormalized by electromagnetic field fluctuations in
the case of a two-component superconductor on a triangular lattice
in Sec. \ref{s:triangular}. In Sec. \ref{s:tetragonal}, we repeat
the same procedure for the case of a tetragonal lattice. In Sec. \ref{sec:conclusions},
we summarize and discuss our results, presenting our concluding remarks.
Appendix \ref{app:taylor} presents additional details of the derivation
of the renormalized free-energy.

\section{\label{s:triangular}Two-component superconductor on the triangular
lattice}

We first consider a two-component unconventional superconductor on
a lattice with threefold rotational symmetry in the presence of electromagnetic
field fluctuations. This applies to the cases of twisted bilayer graphene, with a triangular
moir\'{e} lattice and point group $D_{6}$, and to doped Bi$_{2}$Se$_{3}$
with a trigonal lattice and point group $D_{3d}$. Both of these groups
admit two two-dimensional irreps corresponding to $p_{x}/p_{y}$-wave
or $d_{x^{2}-y^{2}}/d_{xy}$-wave superconducting states -- respectively,
$E_{1}$ and $E_{2}$ in the case of $D_{6}$ and $E_{u}$ and $E_{g}$
in the case of $D_{3d}$. In all these cases, we parametrize the two-component
superconducting order parameter $\psi$ as \cite{Fernandes2019}:
\begin{equation}
\psi=|\psi|e^{i\varphi}\left(\cos\alpha,e^{i\beta}\sin\alpha\right)\mbox{,}\label{eq:parametrization}
\end{equation}
where $\alpha\in[0,2\pi)$ and $\beta\in\left[-\tfrac{\pi}{2},\tfrac{\pi}{2}\right]$.
The global phase $\varphi$ can take any values in $[0,2\pi)$.

\subsection{Renormalized free-energy functional}

\label{talha2} We now generalize the approach of Ref. \citep{HalperinLubenskyMa1974}
of integrating out the electromagnetic field fluctuations for the
case of a two-component superconductor in a lattice with threefold
rotational symmetry. Denoting by $\mathbf{A}$ the electromagnetic
vector potential, and using the same notation as Ref. \citep{HalperinLubenskyMa1974},
the Ginzburg-Landau free-energy density has the form 
\begin{equation}
\mathcal{F}\left[\psi,\mathbf{A}\right]=\mathcal{F}_{0}\left[\psi\right]+\mathcal{F}_{\mathrm{grad}}\left[\psi,\mathbf{A}\right]+\frac{1}{8\pi\mu_{0}}\left(\grad\times\mathbf{A}\right)^{2}\mbox{,}\label{eq:combinedEnergy}
\end{equation}
where $\mathcal{F}_{\mathrm{0}}\left[\psi\right]$ does not contain
gradients of the superconducting order parameter and $\mathcal{F}_{\mathrm{grad}}\left[\psi,\mathbf{A}\right]$
contains all the symmetry allowed couplings between $\psi$ and $\mathbf{A}$.
The last term is the free massless action of the gauge field. Here,
$\mu_{0}$ is the magnetic permeability. The first term on the right-hand
side of Eq. \eqref{eq:combinedEnergy} is given by \cite{SigristUeda1991,FernandesVenderbos2018,HeckerSchmalian2018}
    \begin{equation}
        \begin{aligned}
            \mathcal{F}_{0} & [\psi]=\frac{r}{2}|\psi|^{2}+\frac{u}{4}|\psi|^{4}+\frac{g}{4}\left[\left(\bar{\psi}\tau_{3}\psi\right)^{2}+\left(\bar{\psi}\tau_{1}\psi\right)^{2}\right]\mbox{,}
        \end{aligned}
        \label{eq:MFenergyTriangular}
    \end{equation}
where $\tau_{i}$ refers to the Pauli matrices acting on the two-dimensional
space of $\psi$ (with $i=1,2,3$ ) and $\bar{\psi}$ is the transposed
complex conjugate of $\psi$. The parameter $r$ changes sign at the
bare transition temperature $T_{0}$ as $r=r_{0}(T-T_{0})/T_{0}$,
with $r_{0}>0$. Moreover, the conditions $u>0$ and $g+u>0$ must
hold for $\mathcal{F}_{0}[\psi]$ to be bounded from below. In terms
of the parametrization (\ref{eq:parametrization}), we have:

    \begin{equation}
        \mathcal{F}_{0}[\psi]=\frac{r}{2}|\psi|^{2}+\frac{u}{4}|\psi|^{4}+\frac{g}{4}|\psi|^{4}\left(\sin^{2}2\alpha\cos^{2}\beta+\cos^{2}2\alpha\right).
        \label{eq:MFenergyTriangular2}
    \end{equation}

\begin{figure}
\includegraphics{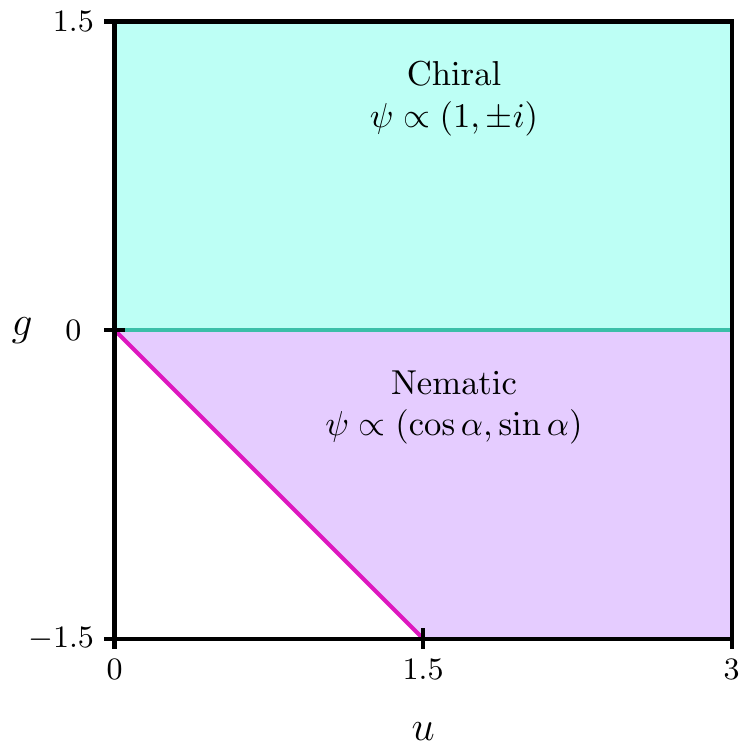} \caption{\label{fig:MFphaseDiagramTriangular}Mean-field phase diagram, in
the $(u,g)$ parameter-space, for a two-component superconductor
on a triangular lattice, based on the Landau free-energy expansion
to quartic order in $\psi$ shown in Eq. \eqref{eq:MFenergyTriangular2}.
The white area in this plot corresponds to the regions in parameter-space
where the free energy in Eq.  \eqref{eq:MFenergyTriangular2} is unbounded.}
\end{figure}

To set the stage, we first review the mean-field results for the case in which gradient terms are absent -- see, e.g., Ref. \cite{SigristUeda1991}. 
Minimizing $\mathcal{F}_{0}[\psi]$, the leading superconducting instabilities
of Eq. \eqref{eq:MFenergyTriangular2} are either the nematic or the
chiral state, both of which onset at $r<0$. Specifically when $g<0$, the leading superconducting state is nematic
and the order parameter has the form $\psi\propto(\cos{\alpha},\sin{\alpha})$
with $\alpha\in[0,2\pi)$. When $g>0$, the leading superconducting
state is chiral and $\psi\propto(1,\pm i)$. The mean-field phase
diagram obtained from minimizing the free-energy in Eq. \eqref{eq:MFenergyTriangular2}
is shown in Fig. \ref{fig:MFphaseDiagramTriangular}. To this order
in $\psi$, the Landau free-energy does not fix $\alpha$ to any particular
value when the nematic state is the minimum. As we will discuss later,
this continuous symmetry is lifted by sixth-order terms in the free-energy.
For simplicity, here we neglect such sixth-order terms, since the
quartic terms are enough to select between the nematic or the chiral
state. In Sec. \ref{s:sixthOrderTerms} we discuss the role of the
sixth-order terms in $\mathcal{F}_{\mathrm{0}}[\psi]$. 

The second term on the right-hand side of Eq. \eqref{eq:combinedEnergy}
consists of a sum of all symmetry allowed gradient terms that couple
$\psi$ and $\mathbf{A}$ \cite{SigristUeda1991}:
\begin{equation}
\begin{aligned} & \mathcal{F}_{\mathrm{grad}}[\psi,\mathbf{A}]=\\
 & K_{1}\left|D_{x}\psi_{1}+D_{y}\psi_{2}\right|^{2}+K_{2}\left|D_{x}\psi_{2}-D_{y}\psi_{1}\right|^{2}\\
 & +K_{3}\left(\left|D_{x}\psi_{1}-D_{y}\psi_{2}\right|^{2}+\left|D_{x}\psi_{2}+D_{y}\psi_{1}\right|^{2}\right)\\
 & +K_{4}\left(\left|D_{z}\psi_{1}\right|^{2}+\left|D_{z}\psi_{2}\right|^{2}\right)\mbox{,}
\end{aligned}
\label{eq:gradTermsTriangular}
\end{equation}
where $D_{x}=\partial_{x}-iq_{0}A_{x}$, etc. are the covariant derivatives
and $q_{0}=2e/\hbar c$. The above $K_{i}$ parameters, known as stiffness
coefficients, penalize spatial variations of the field in different
directions. Importantly, the in-plane stiffness of the order parameter
is not isotropic. We consider the situation in which the order parameter
varies weakly in space whereas the electromagnetic fields vary more
strongly. In this case, we can set $\grad\psi=0$ in the expression
above. This step is formally only justified for type-I superconductors,
as explained in \cite{HalperinLubenskyMa1974}. We will revisit this
assumption in Sec. \ref{sec:conclusions}. With this assumption, the
gradient terms simplify to:
\begin{equation}
\begin{aligned}\mathcal{F}_{\mathrm{grad}}[\psi,\mathbf{A}]= & q_{0}^{2}\gamma_{0}\left|\psi\right|^{2}\left(\!A_{x}^{2}\!+\!A_{y}^{2}\right)\!+\!q_{0}^{2}\gamma_{3}(\bar{\psi}\tau_{3}\psi)\left(\!A_{x}^{2}\!-\!A_{y}^{2}\right)\\
 & +\!2A_{x}A_{y}q_{0}^{2}\gamma_{3}(\bar{\psi}\tau_{1}\psi)\!+\!A_{z}^{2}q_{0}^{2}\gamma_{z}\left|\psi\right|^{2}\mbox{,}
\end{aligned}
\label{eq:GradientTermsTriangular}
\end{equation}
where we have defined the effective stiffness coefficients
\begin{equation}
\gamma_{z}=K_{4}\mbox{,}\quad\gamma_{0}=\frac{K_{1}+K_{2}+2K_{3}}{2}\quad\mbox{and}\quad\gamma_{3}=\frac{K_{1}-K_{2}}{2}.
\end{equation}
In a layered quasi two-dimensional system, the magnitude of $\gamma_{z}$
should be much smaller than that of $\gamma_{0}$.
However, as it will be clear later on, our result is not too sensitive
to variations in $\gamma_{z}$.

To define the effective free-energy density of the single variable $\psi$,
$\mathcal{F}_{\mathrm{eff}}[\psi]$, we take the trace over the physically
allowed dynamic degrees of freedom of $\mathbf{A}$. In other words,
the functional integral that defines $\mathcal{F}_{\mathrm{eff}}[\psi]$
is done over all the purely transverse configurations of the vector
potential, $\mathbf{A}^{\perp}$, 
\begin{equation}
\begin{aligned}e^{-\beta F_{\mathrm{eff}}[\psi]} & =e^{-\beta F_{\mathrm{0}}[\psi]}\\
 & \times\int D\mathbf{A}^{\perp}\,e^{-\beta \int dV \{ \mathcal{F}_{\mathrm{grad}}\left[\psi,\mathbf{A}\right]+\frac{1}{8\pi\mu_{0}}\left(\grad\times\mathbf{A}\right)^{2} \}}\mbox{,}
\end{aligned}
\label{eq:funcint}
\end{equation}
where $\beta=1/\left(k_{B}T\right)$ and $F_i$ denotes the integrated free-energy density $\mathcal{F}_i$. It is convenient to proceed
in the Coulomb gauge, $\nabla\cdot\mathbf{A}=0$, where the Fourier
component of the vector potential that is parallel to the wave vector
$\mathbf{k}$ vanishes, 
\begin{equation}
\mathbf{A}_{\mathbf{k}}\cdot\hat{\mathbf{k}}=0.
\end{equation}

To impose the above condition, we move to the spherical coordinate
system $\mathbf{k}=k\left(\sin\theta\cos\phi,\sin\theta\sin\phi,\cos\theta\right)$
and consider the spherical basis formed by the unit vectors 
\begin{equation}
\begin{aligned}\hat{\mathbf{k}} & =\left(\sin\theta\cos\phi,\sin\theta\sin\phi,\cos\theta\right)\mbox{,}\\
\hat{\boldsymbol{\theta}} & =\left(\cos\theta\cos\phi,\cos\theta\sin\phi,-\sin\theta\right)\,,\\
\hat{\boldsymbol{\phi}} & =\left(-\sin\phi,\cos\phi,0\right)\mbox{.}
\end{aligned}
\end{equation}
In this new basis, the Fourier components $\mathbf{A}_{\mathbf{k}}$
are denoted as 
\begin{equation}
\begin{aligned}\mathbf{A}_{\mathbf{k}} & =\vphantom{\int}A_{k\mathbf{k}}\hat{\mathbf{k}}+\vphantom{\int}A_{\theta\mathbf{k}}\hat{\boldsymbol{\theta}}+A_{\phi\mathbf{k}}\hat{\boldsymbol{\phi}}\mbox{, }\end{aligned}
\end{equation}
in terms of which the transverse component of the electromagnetic
field becomes simply $\mathbf{A}_{\mathbf{k}}^{\perp}=\left(A_{\theta\mathbf{k}},A_{\phi\mathbf{k}}\right).$
Thus, in the Cartesian basis, the Fourier components $\mathbf{A}_{\mathbf{k}}$
are given by: 
\begin{equation}
\begin{aligned}\begin{pmatrix}A_{x\mathbf{k}}\\
A_{y\mathbf{k}}\\
A_{z\mathbf{k}}
\end{pmatrix}=\begin{pmatrix}\cos\theta\cos\phi A_{\theta\mathbf{k}}-\sin\phi A_{\phi\mathbf{k}}\\
\cos\theta\sin\phi A_{\theta\mathbf{k}}+\cos\phi A_{\phi\mathbf{k}}\\
-\sin\theta A_{\theta\mathbf{k}}
\end{pmatrix}\mbox{,}\end{aligned}
\label{eq:basischange}
\end{equation}

As a result, Eq. \eqref{eq:funcint} can be written in terms of $A_{\theta\mathbf{k}}$
and $A_{\phi\mathbf{k}}$ as
\begin{equation}
\begin{aligned}e^{-\beta F_{\mathrm{eff}}[\psi]} & =e^{-\beta F_{\mathrm{0}}[\psi]}\int D\mathbf{A}^{\perp}\,e^{-\tfrac{\beta\psi_{s}^{2}}{8\pi\mu_{0}}\int d^{3}k\,\mathbf{A}_{\mathbf{k}}^{\perp}\mathbf{M}_{\mathbf{k}}\mathbf{A}_{\mathbf{k}}^{\perp T}}\mbox{,}\end{aligned}
\label{eq:funcint2}
\end{equation}
where we have defined $\psi_{s}^{2}=8\pi\gamma_{0}q_{0}^{2}\mu_{0}|\psi|^{2}$ and $\mathbf{M}_{\mathbf{k}}$ is a $2\times2$ matrix with components 
\begin{widetext}
\begin{equation}
\begin{aligned}\left(\mathbf{M}_{\mathbf{k}}\right)_{\theta\theta} & =\cos^{2}\!\theta\left(1\!-\!\tfrac{\gamma_{z}}{\gamma_{0}}\!+\!\tfrac{\gamma_{3}}{\gamma_{0}}\cos2\alpha\cos2\phi\!+\!\tfrac{\gamma_{3}}{\gamma_{0}}\cos\beta\sin2\alpha\sin2\phi\right)\!+\!\tfrac{\gamma_{z}}{\gamma_{0}}\!+\!\tfrac{k^{2}}{\psi_{s}^{2}}\:\mbox{,}\\
\left(\mathbf{M}_{\mathbf{k}}\right)_{\phi\phi} & =1-\tfrac{\gamma_{3}}{\gamma_{0}}\left(\cos2\alpha\cos2\phi+\cos\beta\sin2\alpha\sin2\phi\right)+\tfrac{k^{2}}{\psi_{s}^{2}}\:\mbox{,}\\
\left(\mathbf{M}_{\mathbf{k}}\right)_{\theta\phi} & =\left(\mathbf{M}_{\mathbf{k}}\right)_{\phi\theta}=\tfrac{\gamma_{3}}{\gamma_{0}}\cos\theta\left(\cos\beta\sin2\alpha\cos2\phi-\cos2\alpha\sin2\phi\right)\:\mbox{.}
\end{aligned}
\end{equation}
\end{widetext}

Thus, $\tfrac{\beta\psi_{s}^{2}}{8\pi\mu_{0}}\mathbf{M}_{\mathbf{k}}$
is the ``mass matrix'' of the gauge field. Above the superconducting
transition, where the superconducting order parameter $\psi_{s}$
is zero, the mass matrix has zero determinant, indicative of a massless
field. For a non-zero $\psi_{s}$, the functional integral in Eq.
\eqref{eq:funcint2} only converges if both eigenvalues of the matrix
$\mathbf{M}_{\mathbf{k}}$ are positive, i.e., if the gauge field
becomes massive. The conditions for this to happen are that both $\gamma_{0}$
and $\gamma_{z}$ should be positive and $|\gamma_{3}|<\gamma_{0}$.

The result of the functional integration over all physical configurations
of $\mathbf{A}$ gives the effective free-energy density functional for $\psi$,
which is a sum of two terms 
\begin{equation}
\begin{aligned}\mathcal{F}_{\mathrm{eff}}\left[\psi\right] & =\mathcal{F}_{\mathrm{0}}\left[\psi\right]+\mathcal{F}_{\mathrm{EM}}[\psi]\mbox{.}\end{aligned}
\label{eq:effEnergyTriangular}
\end{equation}
The first term, $\mathcal{F}_{\mathrm{0}}[\psi]$, was defined in
Eqs. \eqref{eq:MFenergyTriangular} or \eqref{eq:MFenergyTriangular2} whereas the second term $\mathcal{F}_{\mathrm{EM}}[\psi]$
is given by the result of the Gaussian integration over the electromagnetic
fields: 
\begin{equation}
\begin{aligned} & \mathcal{F}_{\mathrm{EM}}[\psi]=\frac{4T\Lambda^{3}}{3(2\pi)^{2}}\ln\left(\psi_{s}\right)\\
 & +\frac{T\psi_{s}^{3}}{2(2\pi)^{3}}\int_{0}^{2\pi}d\phi\int_{-1}^{1}dx\int_{0}^{\tfrac{\Lambda}{\psi_{s}}}dq\,q^{2}\ln\left(c+bq^{2}+q^{4}\right)\mbox{.}
\end{aligned}
\label{eq:EMfluctuationsTerm}
\end{equation}
In Eq. \eqref{eq:EMfluctuationsTerm}, we performed a change
of variables to $x=\cos\theta$ and $q=k/\psi_{s}$. Here, $\Lambda$
is the momentum cutoff and the polynomial $c+bq^{2}+q^{4}=\det\mathbf{M}_{\mathbf{k}}$.
The dimensionless quantities $b$ and $c$ are given by 
\begin{equation}
\begin{aligned}\!b= & \frac{\gamma_{z}}{\gamma_{0}}+1+\left(1-\frac{\gamma_{z}}{\gamma_{0}}\right)x^{2}\\
 & -\frac{\gamma_{3}}{\gamma_{0}}\left(1-x^{2}\right)\left(\cos2\alpha\cos2\phi+\cos\beta\sin2\alpha\sin2\phi\right)\,,\\
\!c= & \frac{\gamma_{z}}{\gamma_{0}}\!+\!\left[1-\frac{\gamma_{z}}{\gamma_{0}}-\left(\frac{\gamma_{3}}{\gamma_{0}}\right)^{2}\left(\cos^{2}\beta\sin^{2}2\alpha+\cos^{2}2\alpha\right)\right]\!x^{2}\!\\
 & -\frac{\gamma_{z}\gamma_{3}}{\gamma_{0}^{2}}\left(1-x^{2}\right)\left(\cos2\alpha\cos2\phi+\cos\beta\sin2\alpha\sin2\phi\right).
\end{aligned}
\label{eq:bc}
\end{equation}
In order to extract from $\mathcal{F}_{\mathrm{EM}}[\psi]$ the leading
terms in the order parameter, it is necessary to Taylor expand the
logarithm before integrating. After defining 
\begin{equation}
\begin{aligned}a_{\pm}^{2}=\frac{b}{2}\pm\frac{\sqrt{b^{2}-4c}}{2}\end{aligned}
\label{eq:a}
\end{equation}
we rewrite the integral $\mathcal{F}_{\mathrm{EM}}[\psi]$ as an infinite
sum (see Appendix \ref{app:taylor} for details)
\begin{equation}
\begin{aligned} & \mathcal{F}_{\mathrm{EM}}[\psi]=-\frac{T\psi_{s}^{3}}{48\pi^{2}}\int_{0}^{2\pi}d\phi\int_{-1}^{1}dx\left(a_{+}^{3}+a_{-}^{3}\right)\\
 & \!\!\!+\sum_{n=1}^{\infty}\frac{T\psi_{s}^{2n}}{2(2\pi)^{3}}\int_{0}^{2\pi}d\phi\int_{-1}^{1}dx\frac{(-1)^{n-1}\Lambda^{-2n+3}}{n(-2n+3)}\left(a_{+}^{2n}+a_{-}^{2n}\right).
\end{aligned}
\label{eq:Fem_series}
\end{equation}
The series that contains even powers of $\psi_{s}$, i.e. $\psi_{s}^{2n}$,
simply renormalizes the existing analytic terms in the bare Landau
free-energy. For $n>1$, these corrections are small due to the cutoff
pre-factor $\Lambda^{-2n+3}$. For $n=1$, the correction is independent
of the angles $\alpha$ and $\beta$ and results in a renormalization
of the bare transition temperature $T_{0}$. Therefore, hereafter, we ignore the infinite series and 
focus only on the cubic term of Eq. \eqref{eq:Fem_series}:

\begin{equation}
\begin{aligned}\mathcal{F}_{\mathrm{EM}}[\psi]=- & \frac{T\psi_{s}^{3}}{48\pi^{2}}\int_{0}^{2\pi}d\phi\int_{-1}^{1}dx\,\left(a_{+}^{3}+a_{-}^{3}\right).\end{aligned}
\label{eq:Fem}
\end{equation}

The above cubic term is a non-analytic function of $\psi$. Non-analytic
contributions to the Ginzburg-Landau free-energy are generally expected to
arise when a massless field is integrated out -- see for instance
the case of nematic order parameters coupling to acoustic phonon modes \cite{Schmalian2016,Paul2017,Fernandes_Venderbos2020,Hecker2022}.
If we set $\gamma_{3}=0$ and $\gamma_{0}=\gamma_{z}$, it follows
that $b=2$ and $c=1$, such that $a_{+}=a_{-}=1$. In this case,
Eq. \eqref{eq:Fem} gives a cubic term with a negative coefficient,
as in the case of an $s$-wave superconductor \cite{HalperinLubenskyMa1974}.
What makes our case different from the $s$-wave case is the additional
stiffness coefficient $\gamma_{3}$, which is absent for a single-component
superconductor, and which makes $\mathcal{F}_{\mathrm{EM}}[\psi]$
depend on the relative angles $\alpha$ and $\beta$.

We first analyze numerically the dependence of the cubic term $\mathcal{F}_{\mathrm{EM}}[\psi]$
on $\alpha$ and $\beta$. It is convenient to express the cubic term
in terms of the dimensionless integral $f^{(3)}$ that depends only
on the ratios between the stiffness coefficients $\frac{\gamma_{3}}{\gamma_{0}}$
and $\frac{\gamma_{z}}{\gamma_{0}}$ and on the angles $\alpha$ and
$\beta$:
\begin{equation}
\begin{aligned}\mathcal{F}_{\mathrm{EM}}[\psi]=\frac{T\psi_{s}^{3}}{12\pi}\,f^{(3)}\left(\frac{\gamma_{3}}{\gamma_{0}},\frac{\gamma_{z}}{\gamma_{0}},\alpha,\beta\right)\mbox{.}
\end{aligned}
\label{eq:FemNumerical}
\end{equation}
with:

\begin{equation}
f^{(3)}\equiv-\frac{1}{4\pi}\int_{0}^{2\pi}d\phi\int_{-1}^{1}dx\,\left(a_{+}^{3}+a_{-}^{3}\right)\label{eq:Fem_Numerical_aux}
\end{equation}

\begin{figure}
\includegraphics{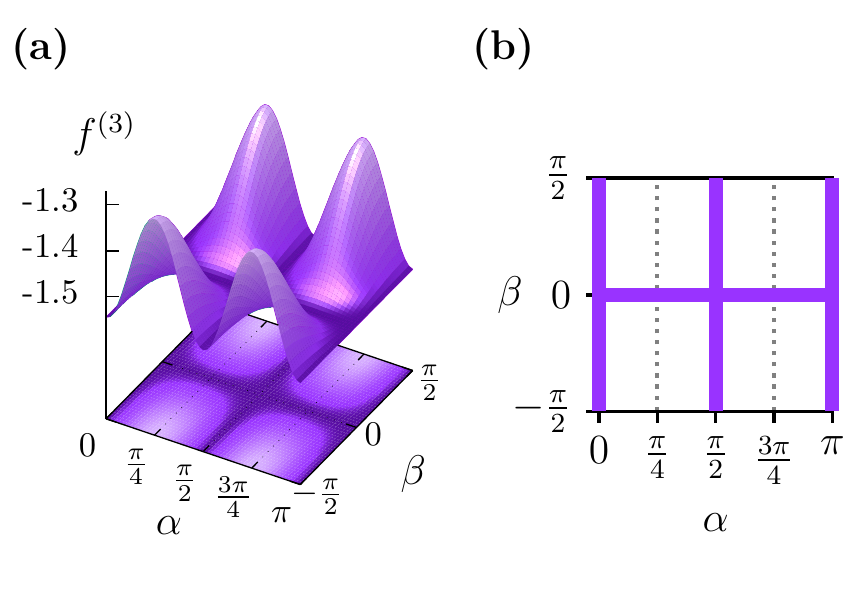} \vspace{-0.75cm}
 \caption{\label{gamma1EQUALgamma3}\textbf{(a)} Plot of $f^{(3)}\left(\tfrac{\gamma_{3}}{\gamma_{0}},\tfrac{\gamma_{z}}{\gamma_{0}},\alpha,\beta\right)$
for fixed $\tfrac{\gamma_{3}}{\gamma_{0}}=0.8$ and $\tfrac{\gamma_{z}}{\gamma_{0}}=0.1$
as a function of $\alpha$ and $\beta$. \textbf{(b)} Location of
the minima on the $(\alpha, \beta)$ plane. The minima correspond
to a nematic state with order parameter $\psi\propto(\cos\alpha,\sin\alpha)$,
where $\alpha$ is not fixed to be any particular value.}
\end{figure}

We analyzed $f^{(3)}$ by plotting it as a function of $\alpha$ and
$\beta$ for varying $\gamma_{z}/\gamma_{0}\in[0,1]$ and $\gamma_{3}/\gamma_{0}\in[-1,1]$.
In all cases we studied, we found $f^{(3)}<0$, like the simpler case
of the $s$-wave superconductor treated in \cite{HalperinLubenskyMa1974}. More importantly, the minima
of $f^{(3)}$ occured for $\beta=0$, with an undefined value of $\alpha$.
This corresponds to a nematic state parametrized by $\psi\propto(\cos\alpha,\sin\alpha)$.
In Fig. \ref{gamma1EQUALgamma3}, we illustrate this behavior by showing
a plot of $f^{(3)}$ for the particular case $\tfrac{\gamma_{z}}{\gamma_{0}}=0.1$
and $\tfrac{\gamma_{3}}{\gamma_{0}}=0.8$. For simplicity, we restrict
$\alpha$ to the range $[0,\pi)$ since the free energy is invariant
under the shift $\alpha\rightarrow\pi+\alpha$.

To gain further insight on these numerical results, we perform an
analytic expansion of $\mathcal{F}_{\mathrm{EM}}[\psi]$ to second
order in $\gamma_{3}/\gamma_{0}$. We find:

\begin{equation}
\begin{aligned} & \mathcal{F}_{\mathrm{EM}}[\psi]\approx-\frac{T\psi_{s}^{3}}{12\pi}\left\{ h_{1}\left(\frac{\gamma_{z}}{\gamma_{0}}\right)+\right.\\
 & \left.+\left(\frac{\gamma_{3}}{\gamma_{0}}\right)^{2}h_{2}\left(\frac{\gamma_{z}}{\gamma_{0}}\right)\left(\sin^{2}2\alpha\cos^{2}\beta+\cos^{2}2\alpha\right)\right\} \mbox{,}
\end{aligned}
\label{eq:cubicafterTaylor}
\end{equation}
where $h_{1}(x)$ and $h_{2}(x)$ are given by 
    \begin{equation}
        \begin{aligned}
            h_{1}(x) &             =\frac{1}{8}\left[10+3x+\frac{3x^{2}}{\sqrt{1-x}}\arctanh\left(\sqrt{1-x}\right)\right]\\
            h_{2}(x) & =\frac{3}{{128(1-x)^{2}}}\left[\right.8x^{2}-5x-6+16(1-x)\ln x\\
            & \quad\left.+\frac{19x^{2}-48x+32}{\sqrt{1-x}}\ln\left(\frac{\sqrt{1-x}+1}{\sqrt{x}}\right)\right]\mbox{.}
        \end{aligned}
        \label{eq:fg}
    \end{equation}
    
\begin{figure}
\includegraphics{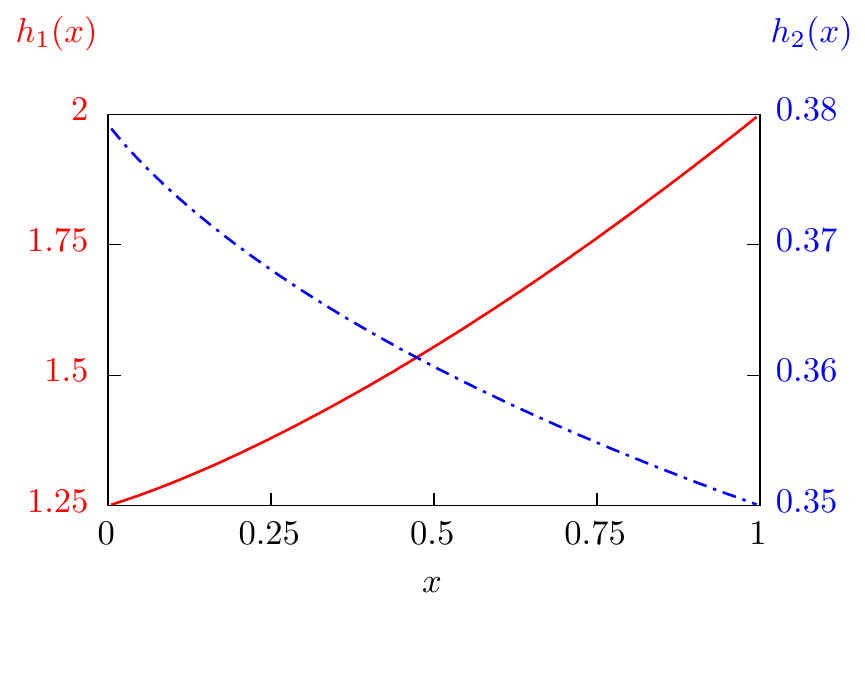} \vspace{-1.3cm}
 \caption{\label{fig:h1h2plots} Plots of the functions $h_{1}(x)$ and $h_{2}(x)$
defined in Eq. \eqref{eq:fg}. Note that both functions are positive
and vary weakly in the interval $0<x<1$.}
\end{figure}

Both $h_{1}(x)$ and $h_{2}(x)$ are plotted in Fig. \ref{fig:h1h2plots}.
We note that the changes in $h_{1}(x)$ and $h_{2}(x)$ in the range
$0<x<1$ are relatively small, implying that our results should not
depend significantly on the value of $\gamma_{z}/\gamma_{0}$. More
importantly, both functions are positive for $0<x<1$, which implies
that the overall coefficient of the cubic term is negative. For later
convenience, we re-express Eq. (\ref{eq:cubicafterTaylor}) as:
    \begin{equation}
        \begin{aligned}
            \mathcal{F}_{\mathrm{EM}} & [\psi]=\\
            & -\frac{\lambda_{0}}{3}|\psi|^{3}\left[1+\frac{\lambda_{3}}{\lambda_{0}}\left(\cos^{2}2\alpha+\sin^{2}2\alpha\cos^{2}\beta\right)\right]\mbox{,}
        \end{aligned}
        \label{eq:ursula1}
    \end{equation}
where the positive parameters $\lambda_{0}$ and $\lambda_{3}$ are
defined as 
    \begin{equation}
        \begin{aligned}
            \lambda_{0} & \equiv\sqrt{32\pi}\left(\gamma_{0}q_{0}^{2}\mu_{0}\right)^{3/2}T\:h_{1}\left(\frac{\gamma_{z}}{\gamma_{0}}\right)\quad,\\
            \lambda_{3} & \equiv\left(\frac{\gamma_{3}}{\gamma_{0}}\right)^{2}\sqrt{32\pi}\left(\gamma_{0}q_{0}^{2}\mu_{0}\right)^{3/2}T\:h_{2}\left(\frac{\gamma_{z}}{\gamma_{0}}\right).
        \end{aligned}
        \label{eq:lambdas}
    \end{equation}

As we pointed out above, while such a negative non-analytic cubic
term also appears in the $s$-wave case and in the isotropic $p$-wave
case \citep{HalperinLubenskyMa1974,LiBelitzToner2009}, the novelty
here is that the non-analytic
contribution also depends on $\alpha$ and $\beta$ due to the in-plane anisotropy of the superconducting stiffness. From Eq. (\ref{eq:ursula1}),
since $\lambda_{3}>0$, it is clear that the term $\mathcal{F}_{\mathrm{EM}}[\psi]$
is minimized for $\beta=0$ and arbitrary $\alpha$, which corresponds
to a nematic superconducting instability, in agreement with our numerical
analysis.

\subsection{Leading instability of the renormalized free-energy }

\label{s:minimaEffEnergyTriangular} Having derived an approximate
analytical expression for $\mathcal{F}_{\mathrm{EM}}[\psi]$, we are
now in position to minimize the full free energy $\mathcal{F}_{\mathrm{eff}}[\psi]$
given by Eq. (\ref{eq:effEnergyTriangular}) to find the leading instability
immediately below the superconducting transition temperature. Using
Eqs. \eqref{eq:MFenergyTriangular2} and \eqref{eq:ursula1}, we obtain: 
    \begin{equation}
        \begin{aligned} 
            & \mathcal{F}_{\mathrm{eff}}[\psi]=\frac{r}{2}|\psi|^{2}-\frac{\lambda_{0}}{3}|\psi|^{3}+\frac{u}{4}|\psi|^{4}\\
            & +\left(\frac{g}{4}|\psi|^{4}-\frac{\lambda_{3}}{3}|\psi|^{3}\right)\left(\cos^{2}2\alpha+\sin^{2}2\alpha\cos^{2}\beta\right).
        \end{aligned}
        \label{eq:effEnergyLambdaParametersTriangular}
    \end{equation}

The key point is that the leading superconducting instability of the
system -- chiral or nematic -- is determined by the competition
between the quartic and cubic terms, which share the same functional
dependence on $\alpha$ and $\beta$. While the cubic term always
favors the nematic phase, the quartic term may favor either the nematic
or the chiral state depending on the sign of $g$, as shown in Fig.
\ref{fig:MFphaseDiagramTriangular} above. 

The presence of a negative cubic term renders the superconducting
transition first-order. As a result, one has to compare the free energies
of the two possible solutions -- nematic ($\beta=0$) and chiral
($\beta=\pm\tfrac{\pi}{2}$, $\alpha=\tfrac{\pi}{4}$). Note that,
because the functional dependence of the renormalized free-energy density
$\mathcal{F}_{\mathrm{eff}}[\psi]$ on $\alpha$ and $\beta$ is the
same as the dependence displayed by the bare free-energy density $\mathcal{F}_{0}[\psi]$,
no additional solutions besides the chiral and nematic ones are expected
to arise from the minimization of the free energy. In either case,
after substituting the appropriate values for the angles, the free
energy acquires the same general form:

    \begin{equation}
        \mathcal{F}_{\mathrm{eff}}^{(\mu)}[\psi]=\frac{r}{2}|\psi|^{2}-\frac{\lambda_{\mu}}{3}|\psi|^{3}+\frac{u_{\mu}}{4}|\psi|^{4}\mbox{,}\label{eq:F_mu}
    \end{equation}
where $\mu$ denotes the nematic ($\mu=\mathrm{nem}$) or the chiral
($\mu=\mathrm{ch}$) solution. We have:
    \begin{equation}\label{eq:F_mu_aux}
        \begin{aligned}[c]
            \lambda_{\mathrm{nem}} & =\lambda_{0}+\lambda_{3}\:;\\
            \lambda_{\mathrm{ch}} & =\lambda_{0}\:;
        \end{aligned}
        \qquad
        \begin{aligned}[c]
            u_{\mathrm{nem}}&=u+g\\
            u_{\mathrm{ch}}&=u
        \end{aligned}
    \end{equation}

It is straightforward to minimize Eq. (\ref{eq:F_mu}) with respect
to $\left|\psi\right|$ and find the condition on the reduced temperature
$r$ for which the minimized free energy becomes smaller than that
of the non-superconducting phase. We find that the first-order transition
for the $\mu$ solution takes place at the reduced temperature $r=r_{\mu}$
given by:
    \begin{equation}
        r_{\mu}=\frac{2\lambda_{\mu}^{2}}{9u_{\mu}}\label{eq:r_mu}.
    \end{equation}
At this transition, the superconducting order parameter jumps according
to:
    \begin{equation}
        \Delta|\psi|_{\mu}=\frac{2\lambda_{\mu}}{3u_{\mu}}\label{eq:Delta_psi_mu}  
    \end{equation}

Therefore, the leading (first-order) superconducting instability is
that whose free energy becomes negative first, i.e. the solution with
the largest $r_{\mu}$ value. Using Eqs. (\ref{eq:F_mu_aux}) and
(\ref{eq:r_mu}), the phase boundary $g^{*}(u)$ between
the chiral and nematic phases in the $(u,g)$ parameter-space is
given implicitly by the condition $r_{\mathrm{ch}}=r_{\mathrm{nem}}$,
from which we derive:
    \begin{equation}
        g^{*}(u)=\left[\left(1+\frac{\lambda_{3}}{\lambda_{0}}\right)^{2}-1\right]u.\label{eq:condition}
    \end{equation}
Note that the chiral solution is the leading instability for $g>g^{*}$
whereas the nematic solution is the leading one for $g<g^{*}$. 

\begin{figure}
\includegraphics{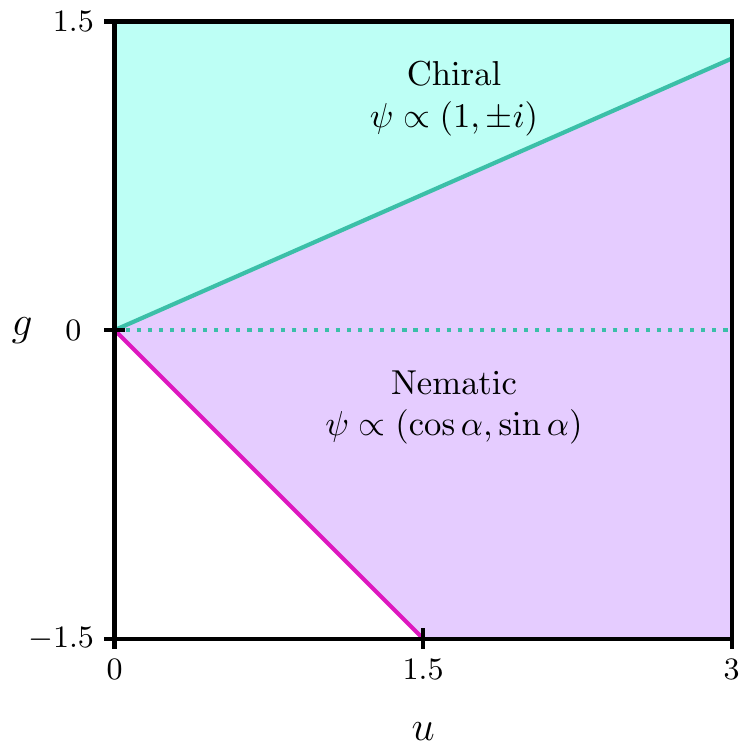} \caption{\label{fig:EffectiveEnergyphaseDiagramTriangular} Phase diagram,
in the $(u,g)$ parameter-space, of the leading superconducting instability
obtained by minimizing the effective free energy in Eq. \eqref{eq:effEnergyLambdaParametersTriangular},
which is renormalized by the electromagnetic field fluctuations. The
dotted line represents the phase boundary of the bare free energy,
see Fig. \ref{fig:MFphaseDiagramTriangular}. The phase boundary separating
the nematic and chiral solutions is a straight line given by Eq. (\ref{eq:condition}).
For this plot, we set $\tfrac{\lambda_{3}}{\lambda_{0}}=0.2$.}
\end{figure}

The phase diagram of the renormalized free-energy is shown
in Fig. \ref{fig:EffectiveEnergyphaseDiagramTriangular}. Compared
with the mean-field phase diagram of the bare free-energy in Fig.
\ref{fig:MFphaseDiagramTriangular}, the main difference is that the
nematic solution becomes the leading instability in a region of the
parameter-space where $g>0$, thus displacing the chiral solution.
Indeed, because $\lambda_{3},\,\lambda_{0}>0$, it follows that $g^{*}>0$.
This implies that the nematic-chiral phase boundary of the renormalized
free-energy moves to the region of the parameter-space where the
chiral solution used to be the leading instability. As a result, the
nematic solution is favored over a wider range of parameters as compared
to the bare free-energy case.

    \begin{figure}
        \includegraphics{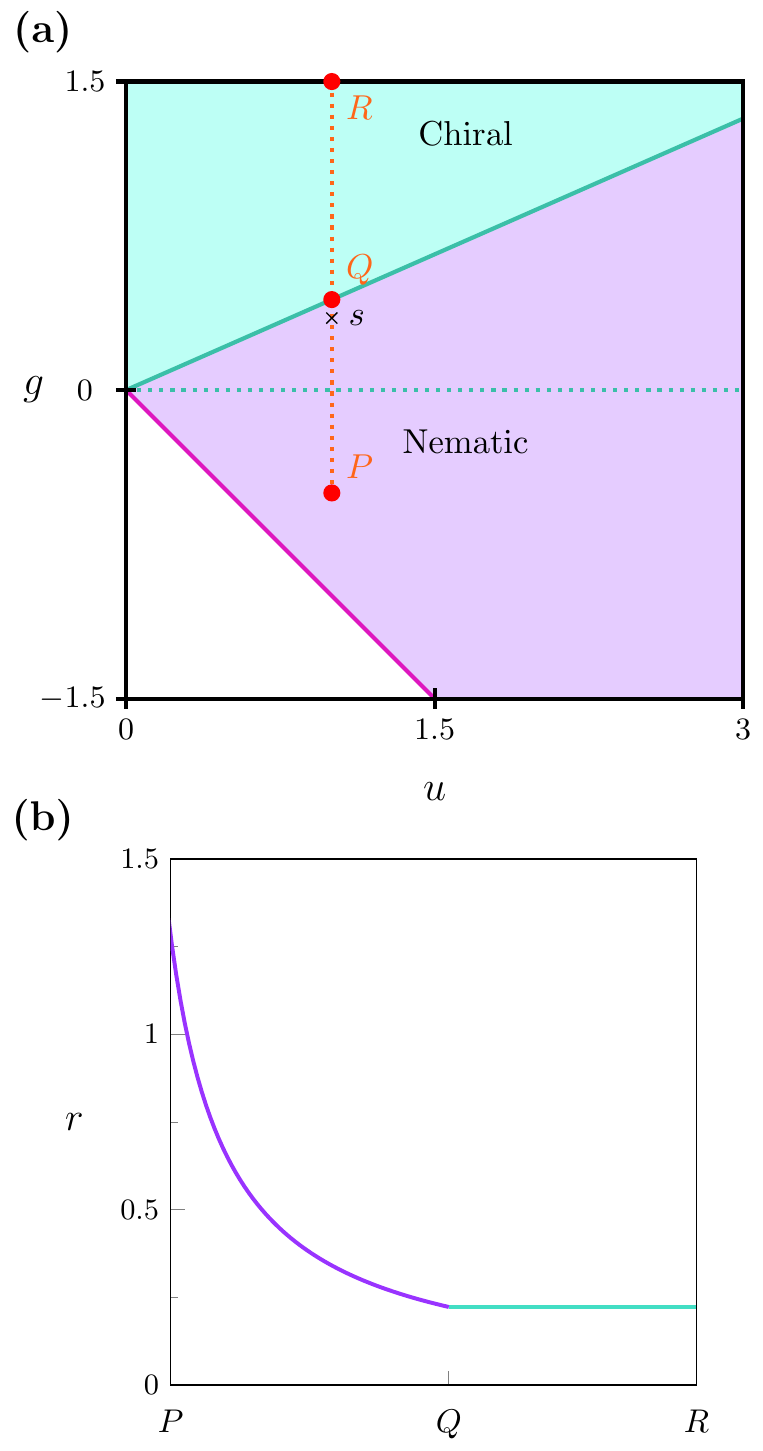} \caption{\label{fig:criticalStuffhex}\textbf{(a)} Definition of the path $P\rightarrow Q\rightarrow R$ in the phase diagram of Fig. \ref{fig:EffectiveEnergyphaseDiagramTriangular}. \textbf{(b)} Evolution of the reduced temperature $r$ where either the nematic (purple line) or the chiral (cyan line) transition takes place, i.e. $\mathrm{max}\left(r_{\mathrm{nem}},\,r_{\mathrm{ch}}\right)$, along the path $P\rightarrow Q\rightarrow R$. For these plots, we used $\lambda_{0}=1   $ and  $\lambda_3=0.2$. We also set $u=1$ along the path $P\rightarrow Q\rightarrow R$.}
    \end{figure}

Another difference between the phase diagrams of Figs. \ref{fig:MFphaseDiagramTriangular}
(bare free-energy) and \ref{fig:EffectiveEnergyphaseDiagramTriangular}
(free-energy renormalized by electromagnetic fluctuations) is that,
in the former, the leading instability is second-order and occurs
always at the reduced temperature $r=0$. In the latter, the transition
is first-order and occurs for a positive $r_{\mu}$ given by Eq. (\ref{eq:r_mu}),
which changes across the phase diagram. This is illustrated in Fig.
\ref{fig:criticalStuffhex}(b), where we plot $\mathrm{max}\left(r_{\mathrm{nem}},\,r_{\mathrm{ch}}\right)$
along the $P-Q-R$ path shown in Fig. 6(a). 

Based on the quantitative estimates of Ref. \citep{HalperinLubenskyMa1974},
one generally expects the cubic coefficients $\lambda_{0}$ and $\lambda_{3}$
to be small, rendering the first-order transition very weak -- in
other words, one expects the jump $\Delta\left|\psi\right|_{\mu}$
in Eq. (\ref{eq:Delta_psi_mu}) to be very small, $\lambda_{i}\ll u$. It
is important to note, however, that this does not imply that the effect
of the electromagnetic field fluctuations on the selection between
the chiral and the nematic phase is negligible. Instead, from the
condition (\ref{eq:condition}), we conclude that this effect is significant
when the ratio between the quartic coefficients $g/u$ is comparable
to the ratio between the cubic coefficients $\lambda_{3}/\lambda_{0}$.
As a result, even though $\lambda_{i}\ll u$, this does not preclude
$g/u\sim\lambda_{3}/\lambda_{0}$.

    \begin{figure*}
        \hspace{-1.7cm} 
        \includegraphics{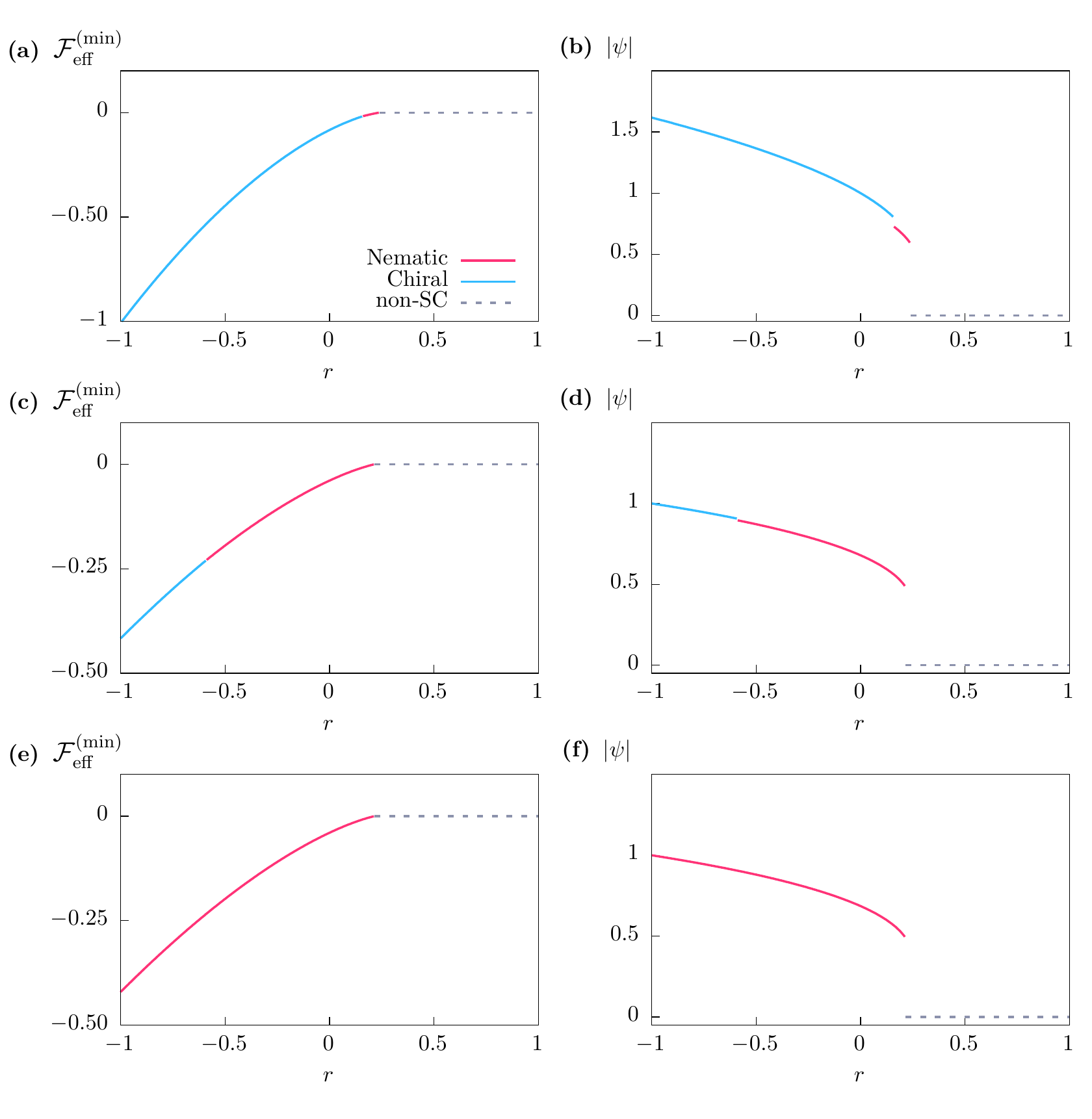} 
        \vspace{-0.7cm}
        \caption{\label{fig:evolutionBelowT1ALL} \textbf{(a)} and \textbf{(b)}: Evolution of the free energy minimum and of the magnitude of the order parameter, as functions of the reduced temperature $r$. Both plots correspond to the point $s$ in the phase diagram of Fig. \ref{fig:criticalStuffhex}(a). 
        This means that we set $u=1$, $g=0.35$, $\lambda_0=1$ and $\lambda_3=0.2$. \textbf{(c)-(f)}: Same as \textbf{(a)}-\textbf{(b)} but in these cases, the sixth-order terms in Eq. (\ref{eq:F_sixth_order}) are included. 
        We set $v_{1}=1$, $v_{2}=0$ and either $v_{3}=0.10$ {[}panels\textbf{(c)} and \textbf{(d)}{]} or $v_{3}=0.15$ {[}panels\textbf{ (e)} and \textbf{(f)}{]}.}
    \end{figure*}

Going back to the effective free-energy in Eq. (\ref{eq:effEnergyLambdaParametersTriangular}),
it is interesting to analyze in more depth the interplay between the
cubic and quartic terms. Naively, one might have expected that the
nematic instability should always be the leading one, since the cubic
term favors the nematic phase, whereas the chiral phase is only favored
by the higher-order quartic term (for $g>0$, of course). The reason
why the quartic term can outcompete the cubic one is because of the
first-order character of the transition. This can be seen by noting
that, immediately below the first-order transition, the combination
$\tilde{g}\equiv\left(\frac{g}{4}-\frac{\lambda_{3}}{3\Delta|\psi|}\right)$
acts as an effective coefficient of the angular-dependent term in
Eq. (\ref{eq:effEnergyLambdaParametersTriangular}), where $\Delta|\psi|$
is the jump in the superconducting order parameter. Plugging in the
value for $\Delta|\psi|_{\mathrm{ch}}$ obtained from Eq. (\ref{eq:Delta_psi_mu}),
we find that $\tilde{g}>0$ in the regime $g>g^{*}$. Clearly, a positive
$\tilde{g}$ favors $\beta=\pm\pi/2$ and $\alpha=\pi/4$, consistent
with a chiral phase. Conversely, substituting the value for $\Delta|\psi|_{\mathrm{nem}}$,
we find that $\tilde{g}<0$ in the regime $g<g^{*}$. A negative effective
coefficient $\tilde{g}$ favors $\beta=0$, consistent with a nematic
phase.

That the nematic phase can be stabilized in a regime where the bare
parameters of the free-energy would predict a chiral phase is the
main result of our paper. Thus, electromagnetic field fluctuations
tilt the balance between the chiral and nematic states in favor of
the latter. Formally, this effect is enabled by the finite stiffness
coefficient $\gamma_{3}$ in Eq. (\ref{eq:GradientTermsTriangular}).
Indeed, $\gamma_{3}=0$ gives $\lambda_{3}=0$, which in turn implies
$g^{*}(u)=0$, recovering the nematic-chiral phase boundary obtained
from the bare free-energy. Note that, as long as the gradient coefficients
$K_{1}$ and $K_{2}$ in Eq. (\ref{eq:gradTermsTriangular}) are different,
$\gamma_{3}$ will be nonzero. Therefore, the microscopic origin of
this effect is the fact that the stiffness of a two-component superconductor
is not isotropic in momentum space.

\subsection{Stability of the superconducting nematic state below $T_{c}$}

\label{s:sixthOrderTerms} The phase diagram obtained in Fig. \ref{fig:EffectiveEnergyphaseDiagramTriangular}
refers to the leading instability immediately below the first-order
transition temperature $T_{c}$ set by $r_{\mathrm{nem}}$ or $r_{\mathrm{ch}}$.
In this section, we investigate the stability of the nematic solution
below the superconducting transition in the region $0<g<g^{*}$. Of
course, since we are employing a Ginzburg-Landau approach, this analysis
is only formally valid near $r_{\mathrm{nem}}$. As such, our calculations
cannot be used to establish what the zero-temperature superconducting
ground state is.

To assess the nematic phase below $r_{\mathrm{nem}}$, it is important
to also include the sixth-order terms of the Landau free-energy $\mathcal{F}_{0}\left[\psi\right]$
that we have neglected so far. This is because, as discussed above,
minimization of the quartic-order free-energy does not fix the value
of the angle $\alpha$ that characterizes the relative amplitude of
the two components of the gap function in the nematic superconducting
state, $\psi_{\mathrm{nem}}\propto(\cos{\alpha},\sin{\alpha})$. A
sixth-order term lowers this artificial $U(1)$ symmetry to a $Z_{3}$
symmetry, as expected for a lattice with threefold rotational symmetry \cite{FernandesVenderbos2018,HeckerSchmalian2018}. We thus include in our analysis the three sixth-order terms
that are allowed by the threefold rotational symmetry of the lattice \cite{SigristUeda1991}:
    \begin{equation}
        \begin{aligned}
            \bar{\mathcal{F}}_{0}\left[\psi\right]= & \frac{v_{1}}{6}|\psi|^{6}+\frac{v_{2}}{6}|\psi|^{2}\left[|\psi|^{4}-\left(\bar{\psi}\tau_{2}\psi\right)^{2}\right]\\
            & +\frac{v_{3}}{6}\left(\bar{\psi}\tau_{3}\psi\right)\left[\left(\bar{\psi}\tau_{3}\psi\right)^{2}-3\left(\bar{\psi}\tau_{1}\psi\right)^{2}\right]\mbox{,}
        \end{aligned}
        \label{eq:F_sixth_order}
    \end{equation}
where new Landau coefficients $v_{1}$, $v_{2}$ and $v_{3}$ were
introduced. To ensure that the free energy remains bounded, they must
satisfy $v_{1}>0$, $v_{1}+v_{2}>0$ and $v_{1}+v_{2}-|v_{3}|>0$.
The first sixth-order term above, with coefficient $v_{1}$, does
not distinguish between the chiral and the nematic states. The second
sixth-order term, with coefficient $v_{2}$, can be rewritten in terms
of the angles $\alpha$ and $\beta$ as:
    \begin{equation}
        \bar{\mathcal{F}}_{0}^{(2)}\left[\psi\right]=\frac{v_{2}}{6}|\psi|^{6}\left(1-\sin^{2}2\alpha\,\sin^{2}\beta\right).
    \end{equation}
Thus, if $v_{2}>0$, the chiral state is favored by this term, whereas
if $v_{2}<0$, the nematic state is favored. As for the third sixth-order
term, with coefficient $v_{3}$, it can be rewritten as:
    \begin{equation}
        \bar{\mathcal{F}}_{0}^{(3)}\left[\psi\right]=\frac{v_{3}}{6}|\psi|^{6}\cos2\alpha\left(\cos^{2}2\alpha-3\sin^{2}2\alpha\,\cos^{2}\beta\right).\label{eq:F_sixth_order_nematic}
    \end{equation}
    \begin{figure}
        \includegraphics{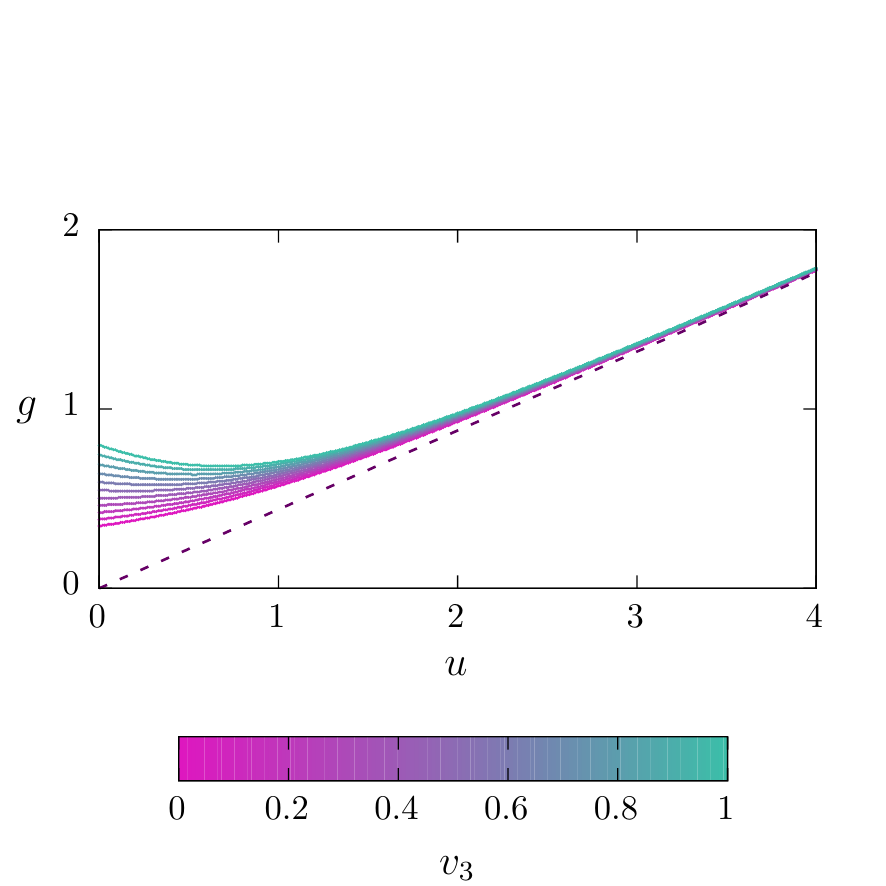} 
        \vspace{-.5cm}
        \caption{\label{fig:boundaries}Nematic-chiral phase boundary in the $(u,g)$ parameter-space. 
        The dashed line is the phase boundary in the absence of the sixth-order terms, as shown in Fig. \ref{fig:EffectiveEnergyphaseDiagramTriangular}.
        Each solid curve represents the phase boundary for different values of $v_{3}$, as shown by the color-scale bar. In all cases, we set $\tfrac{\lambda_{3}}{\lambda_{0}}=0.2$, $v_{1}=1$, and $v_{2}=0$.}
    \end{figure}
This term not only favors the nematic phase ($\beta=0$), regardless
of the sign of $v_{3}$, but it also restricts the allowed values
of $\alpha$ to a discrete set of six values. Indeed, setting $\beta=0$,
we obtain $\bar{\mathcal{F}}_{0}^{(3)}\left[\psi\right]=\tfrac{v_{3}}{6}|\psi|^{6}\cos6\alpha$. As a result, if $v_{3}>0$, this term is minimized by $\alpha=\tfrac{\left(2n+1\right)\pi}{6}$
with $n=0,1,...,5$; conversely, if $v_{3}<0$, minimization gives
$\alpha=\tfrac{2n\pi}{6}$ with $n=0,1,...,5$.

    \begin{figure*}
        \includegraphics{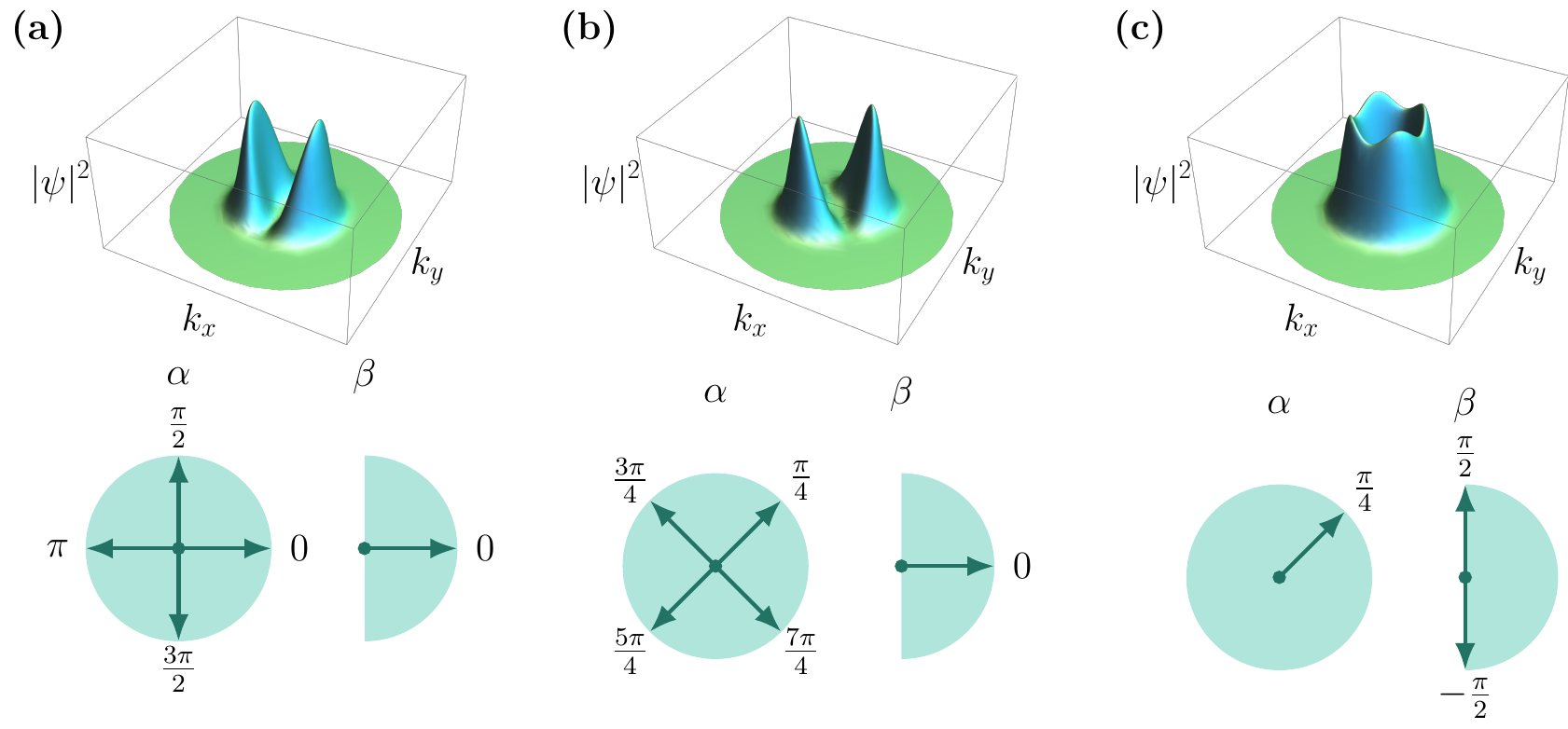} 
        \vspace{-0.4cm}
        \caption{\label{fig:gapsTetra} Absolute value squared of the gap function along a circular Fermi surface for a $(p_{x},\,p_{y})$-wave superconductor on a tetragonal lattice. Three ground states are possible. A similar analysis was shown previously in Ref. \cite{Fernandes2019}. \textbf{(a)} The $\mathrm{B_{1g}}$ nematic superconducting state, which reduces the $C_{4z}$ rotational symmetry of the lattice to a $C_{2z}$ symmetry. In this plot, we set $\alpha=0$. \textbf{(b)} The $\mathrm{B_{2g}}$ nematic superconducting state breaks the $C_{4z}$ rotational symmetry of the lattice as well; in this plot, we used $\alpha=\tfrac{\pi}{4}$. \textbf{(c)} The chiral superconducting state is characterized by $\alpha=\tfrac{\pi}{4}$ and $\beta=\pm\tfrac{\pi}{2}$. It does not break any lattice symmetry, but it breaks time reversal symmetry.}
    \end{figure*}

To investigate the stability of the nematic phase below the superconducting
transition, we numerically minimize the full free-energy $\bar{\mathcal{F}}_{\mathrm{eff}}\equiv\mathcal{F}_{\mathrm{eff}}+\bar{\mathcal{F}}_{0}$,
as given by Eqs. (\ref{eq:effEnergyLambdaParametersTriangular}) and
(\ref{eq:F_sixth_order}), in both the nematic and chiral channels for
$r<r_{\mathrm{nem}}$. Our interest is in the region $0<g<g^{*}$,
where the electromagnetic field fluctuations change the leading instability
from chiral to nematic. For concreteness, we consider the point $s$
in the phase diagram of Fig. \ref{fig:criticalStuffhex}(a), which is
close to the nematic-chiral phase boundary. The evolution of the free
energy minimum, $\mathcal{\bar{F}}_{\mathrm{eff}}^{\mathrm{(min)}}$,
as function of $r$ is shown in Fig. \ref{fig:evolutionBelowT1ALL}
(left panels), accompanied by the evolution of the absolute value of
the superconducting order parameter $\left|\psi\right|$ (right panels).
Without the sixth-order terms {[}panels (a)-(b){]}, the nematic state
undergoes a first-order transition to the chiral state relatively
close to $r_{\mathrm{nem}}$. 

However, upon inclusion of the sixth-order contributions -- particularly
the $\bar{\mathcal{F}}_{0}^{(3)}\left[\psi\right]$ term that is responsible
for enforcing the discreteness of the $\alpha$ values -- we find
that the nematic solution remains the global energy minimum over a
significantly wider range of reduced temperatures $r$ {[}panels (c)-(f){]}.
Interestingly, this effect is apparent even for $\left|v_{3}\right|\ll v_{1}$.
A finite $v_{2}$ can either extend the nematic solution to an even
larger range of reduced temperatures, if $v_{2}<0$, or compress it
to a narrower range, if $v_{2}>0$. Therefore, we conclude that the
sixth-order term \eqref{eq:F_sixth_order_nematic} is important not
only to lift the accidental $U(1)$ symmetry of $\alpha$, but also
to stabilize the nematic phase promoted by the electromagnetic field
fluctuations below the superconducting transition.

Another effect caused by the the sixth-order terms is a change in
the nematic-chiral phase boundary of Fig. \ref{fig:EffectiveEnergyphaseDiagramTriangular}.
As shown in Fig. \ref{fig:boundaries}, upon increasing the coefficient
$v_{3}$ (while keeping $v_{1}$ and $v_{2}$ fixed), the phase boundary
acquires a curvature and is no longer linear. Importantly, this effect
is only significant close to the origin of the $(u,g)$ parameter-space.
As one moves away from the origin, all the boundaries become asymptotically
close to the linear boundary whose slope is determined solely by the
cubic coefficients $\lambda_{0}$ and $\lambda_{3}$. 

\section{Two-component superconductor on the tetragonal lattice}

\label{s:tetragonal} The main result derived in Sec. \ref{s:triangular}
-- that electromagnetic gauge-field fluctuations favor a nematic
over a chiral superconducting state -- is not unique to the triangular
lattice. In this section, we extend the analysis to the case of a
two-component superconductor on a tetragonal lattice. For concreteness,
we consider the point group $D_{4h}$, such that $\psi=\left(\psi_{1},\,\psi_{2}\right)$
can transform as either the $E_{g}$ irrep -- which corresponds to
a $(d_{xz},\,d_{yz})$-wave superconductor -- or the $E_{u}$ irrep
-- corresponding to a $(p_{x},\,p_{y})$-wave superconductor. To start, we review the known results for the mean-field phase diagram (which can be found e.g. in Refs. \cite{SigristUeda1991,Berg2016}), following the notation of Ref. \cite{Fernandes2019}. The
superconducting order parameter can still be parametrized as in Eq.
(\ref{eq:parametrization}). However, instead of two, there are three
possible superconducting ground states: the $\mathrm{B_{1g}}$ nematic
state $\psi=\left(1,\,0\right)/(0,1)$, corresponding to $\alpha=\tfrac{2n\pi}{4}$
and $\beta=0$, with $n=0,...,3$; the $\mathrm{B_{2g}}$ nematic
state $\psi=\left(1,\,\pm1\right)$, corresponding to $\alpha=\tfrac{(2n+1)\pi}{4}$
and $\beta=0$, with $n=0,...,3$; and the chiral state $\psi=\left(1,\,\pm i\right)$,
corresponding to $\alpha=\tfrac{\pi}{4}$, and $\beta=\pm\tfrac{\pi}{2}$.
The corresponding absolute values of the gap function are shown in
Fig. \ref{fig:gapsTetra} for the particular case of a $(p_{x},\,p_{y})$-wave
state -- see also Ref. \cite{Fernandes2019}, where a similar analysis was presented. Both $\mathrm{B_{1g}}$ and $\mathrm{B_{2g}}$ nematic superconducting
states break the fourfold ($C_{4z}$) rotational symmetry of the system,
lowering it to twofold ($C_{2z}$). However, they are not symmetry-equivalent,
as the $\mathrm{B_{1g}}$ state preserves the $\sigma_{v}$ mirror
reflections, whereas the $\mathrm{B_{2g}}$ state preserves the $\sigma_{d}$
mirror reflections.

\begin{figure}
\includegraphics{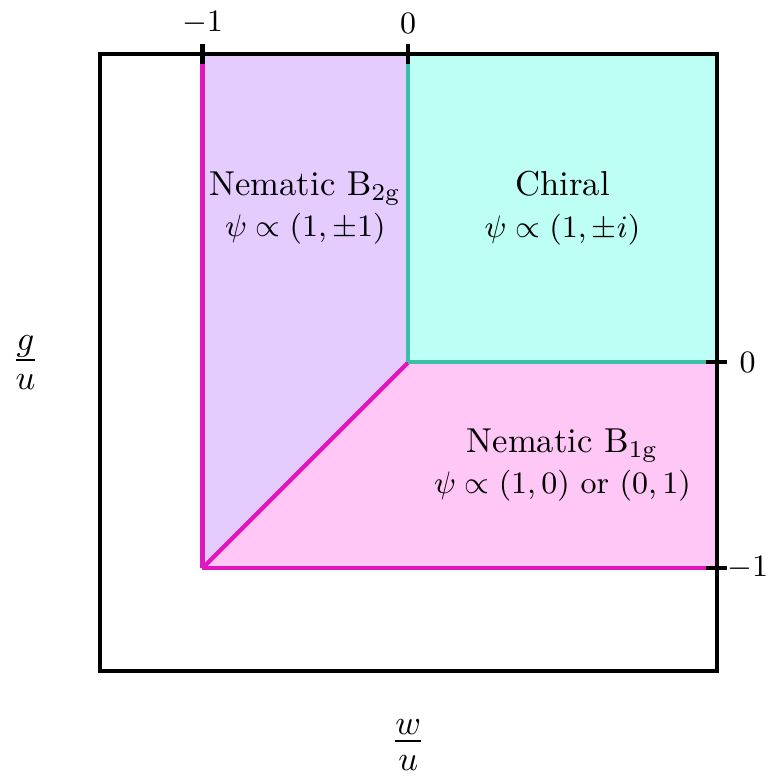} \caption{\label{fig:MFphaseDiagramTetragonal}Mean-field phase diagram in the
$\left(\frac{w}{u},\,\frac{g}{u}\right)$ parameter-space for a two-component
superconductor on a tetragonal lattice, obtained by minimizing the
free-energy in Eq. \eqref{eq:MFenergyTetragonal}. The white area
in this plot corresponds to the parameter-space region where the free-energy
is unbounded from below.}
\end{figure}

To proceed, we write the full Ginzburg-Landau free-energy density as in Eq.
\eqref{eq:combinedEnergy}. The non-gradient terms are given by \cite{SigristUeda1991,Berg2016}:
    \begin{align}
        \begin{aligned}
            \mathcal{F}_{0}[\psi]
        \end{aligned}
            &=\frac{r}{2}|\psi|^{2}+\frac{u}{4}|\psi|^{4}+\frac{g}{4}\left(\bar{\psi}\tau_{3}\psi\right)^{2}+\frac{w}{4}\left(\bar{\psi}\tau_{1}\psi\right)^{2}\label{eq:MFenergyTetragonal}\\
            &=\frac{r}{2}|\psi|^{2}+\frac{u}{4}|\psi|^{4}\nonumber \\
            & +\frac{g}{4}|\psi|^{4}\cos^{2}2\alpha+\frac{w}{4}|\psi|^{4}\sin^{2}2\alpha\cos^{2}\beta.\nonumber 
        \end{align}
In order for $\mathcal{F}_{0}[\psi]$ to be bounded, the Landau parameters
must satisfy the conditions $u>0$, $w+u>0$ and $g+u>0$. Minimization
of the free energy leads to the three possible superconducting solutions
mentioned above. As shown in the mean-field phase diagram of Fig.
\ref{fig:MFphaseDiagramTetragonal}, when $g<\min\left\lbrace 0,w\right\rbrace $,
the leading instability below $r<0$ is the $\mathrm{B_{1g}}$ nematic
superconducting state. When $g>w$ and $w<0$, the selected state
is the $\mathrm{B_{2g}}$ nematic, whereas for $g>0$ and $w>0$,
it is the chiral state.

The gradient terms are given by \citep{SigristUeda1991}:
    \begin{equation}
        \begin{aligned} 
            & \qquad\mathcal{F}_{\mathrm{grad}}[\psi]=K_{1}\!\left[\left|D_{x}\psi_{1}\right|^{2}\!+\!\left|D_{y}\psi_{2}\right|^{2}\right]\\
            & +\!K_{2}\!\left[\left|D_{x}\psi_{2}\right|^{2}\!+\!\left|D_{y}\psi_{1}\right|^{2}\right]\!+\!K_{3}\!\left[\left(D_{x}\psi_{1}\right)^{\ast}\left(D_{y}\psi_{2}\right)\!+\!c.c.\right]\\
            & +\!K_{4}\!\left[\left(D_{x}\psi_{2}\right)^{\ast}\left(D_{y}\psi_{1}\right)\!+\!c.c.\right]\!+\!K_{5}\!\left[\left|D_{z}\psi_{1}\right|^{2}\!+\!\left|D_{z}\psi_{2}\right|^{2}\right]\mbox{,}
        \end{aligned}
        \label{eq:gradTermsTetragonal}
    \end{equation}
where, as in Sec. \ref{s:triangular}, $D_{x}=\partial_{x}-iq_{0}A_{x}$,
etc. are the covariant derivatives and $K_{i}$ are the stiffness
coefficients. Assuming that the order parameter is spatially uniform
in the regime where the gauge-field fluctuations are strong, the equation
above is simplified to: 
\begin{equation}
\begin{aligned}\mathcal{F}_{\mathrm{grad}}[\psi]= & q_{0}^{2}\gamma_{0}\left|\psi\right|^{2}\left(A_{x}^{2}+A_{y}^{2}\right)+q_{0}^{2}\gamma_{3}(\bar{\psi}\tau_{3}\psi)\left(A_{x}^{2}-A_{y}^{2}\right)\\
 & +2A_{x}A_{y}q_{0}^{2}\gamma_{1}(\bar{\psi}\tau_{1}\psi)+A_{z}^{2}q_{0}^{2}\gamma_{z}\left|\psi\right|^{2}\mbox{,}
\end{aligned}
\label{eq:GradientTermsTetragonal}
\end{equation}
where we have defined the effective stiffness coefficients as $\gamma_{z}=K_{5}$,
    \begin{equation}
        \begin{aligned}
            \gamma_{1} & =\frac{K_{3}+K_{4}}{2},\!\quad\gamma_{3}=\frac{K_{1}-K_{2}}{2}\quad\!\mbox{and}\!\quad\gamma_{0}=\frac{K_{1}+K_{2}}{2}.
        \end{aligned}
    \end{equation}
Note that, as compared to the triangular-lattice case, there is an
additional stiffness coefficient in the case of the tetragonal lattice,
since $\gamma_{1}\neq\gamma_{3}$. If these two coefficients were
fine-tuned to acquire the same value, one would recover the results
for the triangular lattice.

We now repeat the same steps as in Sec. \ref{s:triangular} to integrate
out the electromagnetic field fluctuations and obtain the effective
free-energy density
    \begin{equation}
        \begin{aligned}
            \mathcal{F}_{\mathrm{eff}}\left[\psi\right] & =\mathcal{F}_{0}\left[\psi\right]+\mathcal{F}_{\mathrm{EM}}[\psi]\mbox{,}
        \end{aligned}
    \end{equation}
with $\mathcal{F}_{\mathrm{0}}[\psi]$ defined in Eq. \eqref{eq:MFenergyTetragonal}.
The term $\mathcal{F}_{\mathrm{EM}}[\psi]$, resulting from the gauge-field
fluctuations, acquires the same form as in Eq. \eqref{eq:Fem}, with
$a_{\pm}^{2}$ still defined by Eq. \eqref{eq:a}, but with the new dimensionless
quantities $b$ and $c$ given by: 
    \begin{equation}
        \begin{aligned} 
            & \!b=\!\frac{\gamma_{z}}{\gamma_{0}}+1+\left(1-\frac{\gamma_{z}}{\gamma_{0}}\right)x^{2}\\
            & -\!\left(1\!-\!x^{2}\right)\!\left(\frac{\gamma_{3}}{\gamma_{0}}\cos2\alpha\cos2\phi+\!\frac{\gamma_{1}}{\gamma_{0}}\cos\beta\sin2\alpha\sin2\phi\right)
        \end{aligned}
        \label{eq:bcTetra}
    \end{equation}
and 
    \begin{equation}
        \begin{aligned} 
            & \!c=\!\frac{\gamma_{z}}{\gamma_{0}}\!+\!\!\left[1\!-\!\frac{\gamma_{z}}{\gamma_{0}}\!-\!\left(\frac{\gamma_{3}}{\gamma_{0}}\cos2\alpha\right)^{2}\!\!-\!\left(\frac{\gamma_{1}}{\gamma_{0}}\cos\beta\sin2\alpha\right)^{2}\right]\!x^{2}\!\\
            & -\frac{\gamma_{z}}{\gamma_{0}}\!\left(1\!-\!x^{2}\right)\!\left(\frac{\gamma_{3}}{\gamma_{0}}\cos2\alpha\cos2\phi+\frac{\gamma_{1}}{\gamma_{0}}\cos\beta\sin2\alpha\sin2\phi\right).
        \end{aligned}
    \end{equation}

We first study numerically the dependence of the cubic term $\mathcal{F}_{\mathrm{EM}}[\psi]$
on $\alpha$ and $\beta$. As in the preceding section, it is convenient
to express it in terms of the dimensionless integral $f^{(3)}$ given
by Eq. (\ref{eq:Fem_Numerical_aux}), such that:
\begin{equation}
\begin{aligned}\mathcal{F}_{\mathrm{EM}}[\psi]=\frac{T\psi_{s}^{3}}{12\pi}\,f^{(3)}\left(\frac{\gamma_{1}}{\gamma_{0}},\frac{\gamma_{3}}{\gamma_{0}},\frac{\gamma_{z}}{\gamma_{0}},\alpha,\beta\right)\mbox{.}\end{aligned}
\label{eq:FemNumericalTetra}
\end{equation}

\begin{figure}
\includegraphics{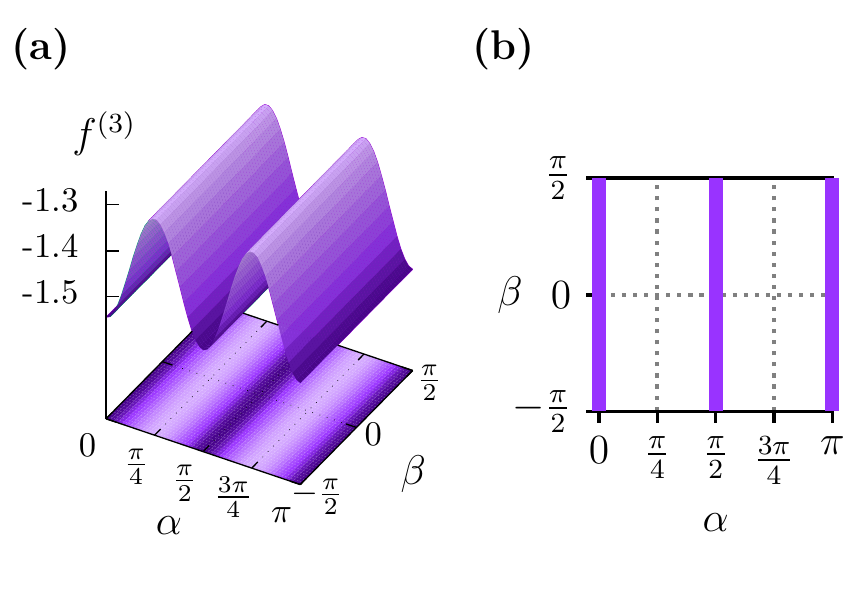} \vspace{-0.75cm}
 \caption{\label{gamma1LESSgamma3}\textbf{(a)} Dimensionless coefficient of
the cubic term $f^{(3)}\left(\alpha,\beta\right)$ as a function of
$\alpha$ and $\beta$ for fixed $\tfrac{\gamma_{1}}{\gamma_{0}}=0$,
$\tfrac{\gamma_{3}}{\gamma_{0}}=0.8$ and $\tfrac{\gamma_{z}}{\gamma_{0}}=0.1$.
\textbf{(b)} Location of the minima of $f^{(3)}\left(\alpha,\beta\right)$
on the $(\alpha,\beta)$ plane. The minima correspond to the $\mathrm{B_{1g}}$
nematic superconducting state with order parameter $\psi\propto(1,0)$
or $\psi\propto(0,1)$.}
\end{figure}

\begin{figure}
\includegraphics{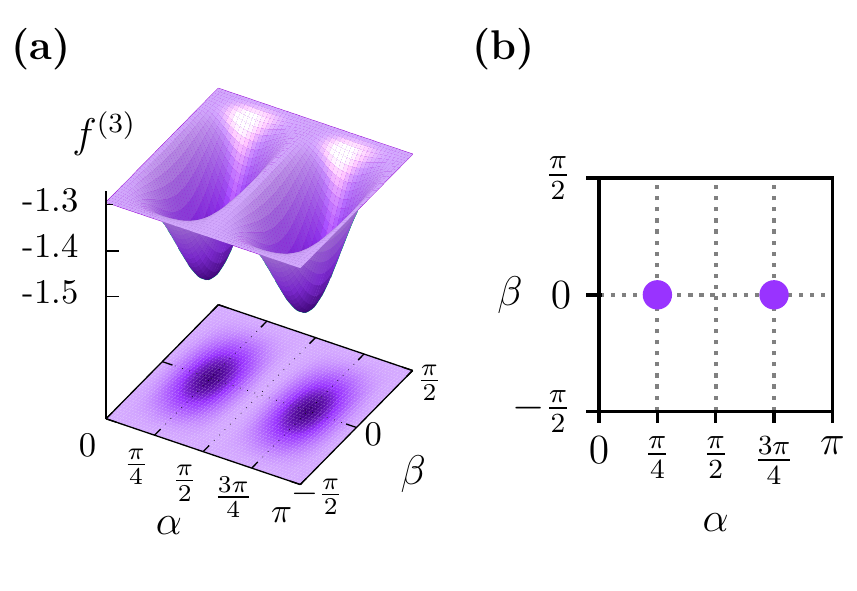} \vspace{-0.75cm}
 \caption{\label{gamma1MOREgamma3}\textbf{(a)} Dimensionless coefficient of
the cubic term $f^{(3)}\left(\alpha,\beta\right)$ as a function of
$\alpha$ and $\beta$ for fixed $\tfrac{\gamma_{1}}{\gamma_{0}}=0.8$,
$\tfrac{\gamma_{3}}{\gamma_{0}}=0$ and $\tfrac{\gamma_{z}}{\gamma_{0}}=0.1$.
\textbf{(b)} Location of the minima of $f^{(3)}\left(\alpha,\beta\right)$
on the $(\alpha,\beta)$ plane. The minima correspond to the $\mathrm{B_{2g}}$
nematic superconducting state with order parameter $\psi\propto(1,\pm1)$.}
\end{figure}

We systematically analyzed $f^{(3)}\left(\alpha,\beta\right)$ numerically
for various values of the stiffness coefficients. Because $f^{(3)}$
only depends on $\cos2\alpha$ and $\sin2\alpha$, we restricted the
range of $\alpha$ values to $[0,\pi]$. The stiffness coefficients
were varied systematically in the ranges $\tfrac{\gamma_{z}}{\gamma_{0}}\in[0,1]$,  $\tfrac{\gamma_{1}}{\gamma_{0}}\in[-1,1]$,
and $\tfrac{\gamma_{3}}{\gamma_{0}}\in[-1,1]$. In all cases,
we found $f^{(3)}\left(\alpha,\beta\right)<0$. More importantly,
the values of $\alpha$ and $\beta$ that minimize $f^{(3)}\left(\alpha,\beta\right)$
were found to always correspond to one of the two nematic superconducting
states. In particular, in the cases where $|\gamma_{1}|<|\gamma_{3}|$,
the minima are located at $\alpha=2n\pi/4$ with integer $n$, corresponding
to the $\mathrm{B_{1g}}$ nematic superconducting state $\psi\propto(1,0)/(0,1)$.
This is illustrated in Fig. \ref{gamma1LESSgamma3}, where we show
$f^{(3)}\left(\alpha,\beta\right)$ for the particular case $\tfrac{\gamma_{1}}{\gamma_{0}}=0$,
$\tfrac{\gamma_{3}}{\gamma_{0}}=0.8$, and $\tfrac{\gamma_{z}}{\gamma_{0}}=0.1$.
Conversely, in all the cases where $|\gamma_{1}|>|\gamma_{3}|$, the
minima are at $\beta=0$ and $\alpha=\left(2n+1\right)\pi/4$ with
integer $n$, corresponding to the $\mathrm{B_{2g}}$ nematic superconducting
state $\psi\propto(1,\pm1)$. Such a behavior is illustrated in Fig.
\ref{gamma1MOREgamma3} for the particular case $\tfrac{\gamma_{1}}{\gamma_{0}}=0.8$, 
$\tfrac{\gamma_{3}}{\gamma_{0}}=0$, and  $\tfrac{\gamma_{z}}{\gamma_{0}}=0.1$.

Following the same steps as in Sec. \ref{s:triangular}, we perform
an analytical expansion of $f^{(3)}\left(\frac{\gamma_{1}}{\gamma_{0}},\frac{\gamma_{3}}{\gamma_{0}},\frac{\gamma_{z}}{\gamma_{0}},\alpha,\beta\right)$
to second order in $\gamma_{1}/\gamma_{0}$ and $\gamma_{3}/\gamma_{0}$.
We obtain:
\begin{equation}
\begin{aligned} & \mathcal{F}_{\mathrm{EM}}[\psi]\sim-\frac{T\psi_{s}^{3}}{12\pi}\left\{ h_{1}\left(\frac{\gamma_{z}}{\gamma_{0}}\right)+\right.\\
 & \left.h_{2}\left(\frac{\gamma_{z}}{\gamma_{0}}\right)\left[\left(\frac{\gamma_{1}}{\gamma_{0}}\sin2\alpha\cos\beta\right)^{2}+\left(\frac{\gamma_{3}}{\gamma_{0}}\cos2\alpha\right)^{2}\right]\right\} \mbox{,}
\end{aligned}
\label{eq:cubicafterTaylorTetra}
\end{equation}
where $h_{1}(x)$ and $h_{2}(x)$ were previously defined in Eq. \eqref{eq:fg}
and plotted in Fig. \ref{fig:h1h2plots}. Minimization of $\mathcal{F}_{\mathrm{EM}}[\psi]$
leads to $\alpha=\left(2n+1\right)\pi/4$ and $\beta=0$ when $\left|\gamma_{1}\right|>\left|\gamma_{3}\right|$
and to $\alpha=2n\pi/4$ when $\left|\gamma_{3}\right|>\left|\gamma_{1}\right|$,
in agreement with the numerical analysis. For convenience, we define
the coefficients: 
    \begin{equation}
        \begin{aligned}
            \lambda_{0} & \equiv\sqrt{32\pi}\left(\gamma_{0}q_{0}^{2}\mu_{0}\right)^{3/2}T\:h_{1}\left(\frac{\gamma_{z}}{\gamma_{0}}\right)\,,\\
            \lambda_{1} & \equiv\sqrt{32\pi}\left(\gamma_{0}q_{0}^{2}\mu_{0}\right)^{3/2}T\:\left(\frac{\gamma_{1}}{\gamma_{0}}\right)^{2}h_{2}\left(\frac{\gamma_{z}}{\gamma_{0}}\right)\mbox{,}\\
            \lambda_{3} & \equiv\lambda_{1}\left(\frac{\gamma_{3}}{\gamma_{1}}\right)^{2},
        \end{aligned}
        \label{eq:lambdas2}
    \end{equation}
and rewrite the cubic term as:
\begin{equation}
\begin{aligned}\mathcal{F}_{\mathrm{EM}} & [\psi]=\\
 & -\frac{\lambda_{0}}{3}|\psi|^{3}\left(1+\frac{\lambda_{1}}{\lambda_{0}}\sin^{2}2\alpha\cos^{2}\beta+\frac{\lambda_{3}}{\lambda_{0}}\cos^{2}2\alpha\right)\mbox{,}
\end{aligned}
\label{eq:ursula2}
\end{equation}

Since $\mathcal{F}_{\mathrm{EM}}[\psi]$ has the same functional dependence
on $\alpha$ and $\beta$ as the bare free-energy density $\begin{aligned}\mathcal{F}_{0}[\psi]\end{aligned}
$, minimization of the total free-energy density $\mathcal{F}_{\mathrm{eff}}\left[\psi\right]$
should yield the same solutions as $\begin{aligned}\mathcal{F}_{0}[\psi]\end{aligned}
$. Like we did in Sec. \ref{s:triangular}, to find the leading instability,
we compare the free energies of the three solutions, since the cubic
term renders the transition first-order. In all cases, after substituting
the values for $\alpha$ and $\beta$ corresponding to each solution,
the free-energy density acquires the same form:
    \begin{equation}
        \mathcal{F}_{\mathrm{eff}}^{(\mu)}[\psi]=\frac{r}{2}|\psi|^{2}-\frac{\lambda_{\mu}}{3}|\psi|^{3}+\frac{u_{\mu}}{4}|\psi|^{4}\label{eq:F_mu_T}
    \end{equation}
where $\mu$ labels the type of solution ($\mu=\mathrm{B_{1g}},\:\mathrm{B_{2g}},\:\mathrm{ch}$)
and:
    \begin{equation}
        \begin{aligned}[c]
            \lambda_{\mathrm{B_{1g}}} & =\lambda_{0}+\lambda_{3}\:;\\
            \lambda_{\mathrm{B_{2g}}} & =\lambda_{0}+\lambda_{1}\:;\\
            \lambda_{\mathrm{ch}} & =\lambda_{0}\:;
        \end{aligned}
        \qquad
        \begin{aligned}[c]
            u_{\mathrm{B_{1g}}}&=u+g\\
            u_{\mathrm{B_{2g}}}&=u+w\\
            u_{\mathrm{ch}}&=u
        \end{aligned}
    \end{equation}
As we showed in Sec. \ref{s:triangular}, the first-order transition
associated with the free-energy in Eq. (\ref{eq:F_mu_T}) takes place
at the reduced temperature $r_{\mu}=\frac{2\lambda_{\mu}^{2}}{9u_{\mu}}$.
Thus, the leading instability is the one with the largest transition
temperature:
    \begin{align}
        r_{\mathrm{B_{1g}}} & =\frac{2\left(\lambda_{0}+\lambda_{3}\right)^{2}}{9\left(u+g\right)}\nonumber \\
        r_{\mathrm{B_{2g}}} & =\frac{2\left(\lambda_{0}+\lambda_{1}\right)^{2}}{9\left(u+w\right)}\nonumber \\
        r_{\mathrm{ch}} & =\frac{2\lambda_{0}^{2}}{9u}
    \end{align}

It is now straightforward to determine the phase boundaries in the
$\left(\frac{w}{u},\,\frac{g}{u}\right)$ parameter-space. The chiral
solution is the leading instability in the region bounded by $w>w^{*}$
and $g>g^{*}$, where: 
    \begin{align}
        \frac{w^{\ast}}{u} & =\left[\left(\frac{\lambda_{1}}{\lambda_{0}}+1\right)^{2}-1\right]\nonumber \\
        \frac{g^{\ast}}{u} & =\left[\left(\frac{\lambda_{3}}{\lambda_{0}}+1\right)^{2}-1\right]\:.\label{eq:wgstar}
    \end{align}
    
    \begin{figure}
        \includegraphics{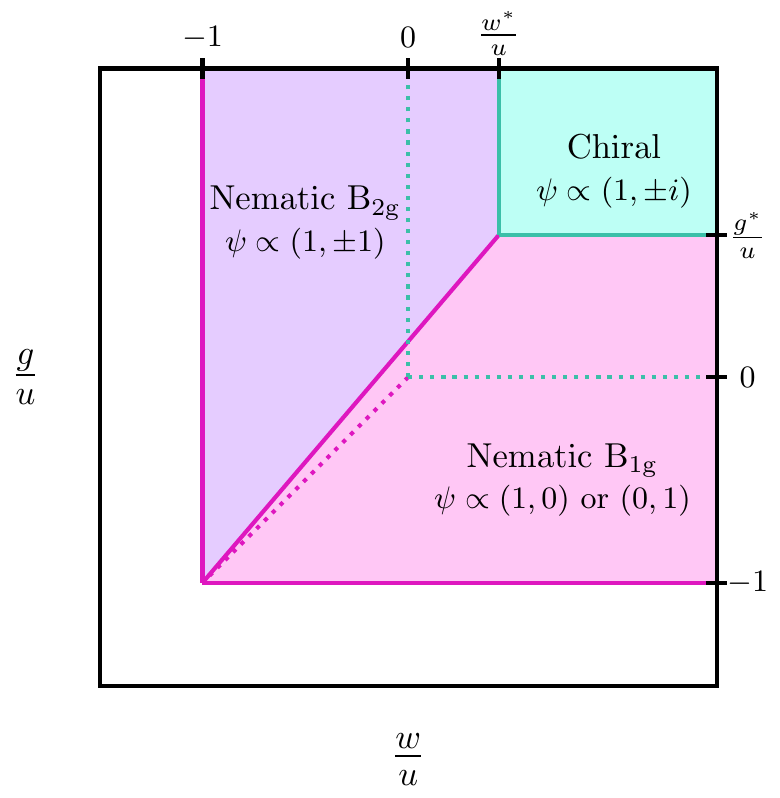} \caption{\label{fig:EffectiveEnergyphaseDiagramTetragonal} Phase diagram, in the $\left(\frac{w}{u},\,\frac{g}{u}\right)$ parameter-space, for a two-component superconductor on a tetragonal lattice obtained from minimization of the free-energy renormalized by electromagnetic field fluctuations. The dotted lines represent the phase boundaries of the mean-field phase diagram (see Fig. \ref{fig:MFphaseDiagramTetragonal}). For this plot, we set $\tfrac{\lambda_{1}}{\lambda_{0}}=0.2$ and $\tfrac{\lambda_{3}}{\lambda_{0}}=0.3$. The  quantities $w^{\ast}$ and $g^{\ast}$ are defined in Eq. \eqref{eq:wgstar}, and in this plot are given by $w^{\ast}=0.44u$ and $g^{\ast}=0.69u$.}
    \end{figure}

The fact that $w^{\ast},g^{\ast}>0$ implies that the region of the phase
diagram where the chiral solution is realized shrinks with respect
to the region occupied by the chiral solution in the mean-field phase
diagram. This is illustrated in Fig. \ref{fig:EffectiveEnergyphaseDiagramTetragonal},
where the renormalized phase boundaries are shown by the solid lines
whereas the bare phase boundaries are given by the dashed lines. Therefore,
after renormalization by the electromagnetic field fluctuations, the
nematic state becomes the leading superconducting instability over
a significant range of parameters for which the mean-field analysis
would predict a chiral state. This result is analogous to the case
of the two-component superconductor on the triangular lattice.

There is, however, one important difference, as there are two symmetry-distinct
nematic superconducting states on the tetragonal lattice, namely the
$\mathrm{B_{1g}}$ and $\mathrm{B_{2g}}$ nematic solutions. Comparing
$r_{\mathrm{B_{1g}}}$ and $r_{\mathrm{B_{2g}}}$, we find that, for
$w<w^{*}$ and $g<g^{*}$, the phase boundary $\tilde{g}\left(w\right)$
separating the two nematic phases is given by:
    \begin{equation}
        \frac{\tilde{g}\left(w\right)}{u}=-1+\left(\frac{w}{u}+1\right)\left(\frac{\lambda_{3}+\lambda_{0}}{\lambda_{1}+\lambda_{0}}\right)^{2}\:,
    \end{equation}
such that the $\mathrm{B_{1g}}$ state is realized for $g<\tilde{g}\left(w\right)$
and the $\mathrm{B_{2g}}$ state, for $g>\tilde{g}\left(w\right)$.
Compared to the phase boundary of the mean-field phase diagram, $\tilde{g}_{\mathrm{MF}}\left(w\right)=w$,
we conclude that, for $\lambda_{3}>\lambda_{1}$, the $\mathrm{B_{1g}}$
nematic solution occupies a region of the parameter-space that was
occupied by the $\mathrm{B_{2g}}$ nematic solution in the mean-field
case. This case is illustrated in Fig. \ref{fig:EffectiveEnergyphaseDiagramTetragonal}.
Conversely, for $\lambda_{1}>\lambda_{3}$, it is the $\mathrm{B_{2g}}$
solution that occupies an expanded region of the parameter-space.

\section{Conclusions}

\label{sec:conclusions} In this paper, we showed that electromagnetic
fluctuations play an important role in the selection between nematic
versus chiral superconductivity for two-component superconductors,
such as $\left(p_{x},p_{y}\right)$-wave and $\left(d_{x^{2}-y^{2}},d_{xy}\right)$-wave
states. Upon integrating out these gauge-field fluctuations, they
generate non-analytic cubic terms in the free-energy that induce a
first-order transition, similarly to the cases of the $s$-wave and
multi-component superconductors with isotropic stiffness analyzed elsewhere \cite{HalperinLubenskyMa1974,Millev1990,LiBelitzToner2009},
as well as of color superconductivity involving quarks and gluons
\cite{Baym2004}. The crucial difference is that, for the two-component superconductors
studied here, the superconducting stiffness -- or, equivalently,
the correlation length -- is not isotropic in the $(x,y)$ plane
due to the crystalline lattice. This makes the non-analytic term in
the free-energy sensitive to whether the superconducting state is
nematic or chiral, favoring the former over the latter. The relevance
of this result stems from the fact that weak-coupling microscopic
calculations generally place the system in a region of the parameter-space
where minimization of the mean-field free-energy predicts a chiral
state. However, as shown here, the non-analytic free-energy term arising
from the gauge-field fluctuations changes the nature of the leading
instability in a significant portion of this parameter-space region
from chiral to nematic. As a result, the effect of the electromagnetic
field fluctuations on the superconducting free-energy provides a mechanism
by which a nematic state can be stabilized over the chiral one, without
requiring fine tuning or coupling to non-superconducting degrees of
freedom.

We emphasize that the size of the effect uncovered here is not necessarily
small, even if the induced transition is very weakly first-order,
as is the case for $s$-wave superconductors. Indeed, a weak first-order
transition generally implies that the coefficients of the cubic terms
($\lambda_{0}$ and $\lambda_{3}$ in our notation for the triangular-lattice
case) are much smaller than the coefficients of the quartic terms
($u$ and $g$ in our notation). However, the change in the leading
superconducting instability from chiral to nematic promoted by the
gauge-field fluctuations takes place when $g/u\lesssim\lambda_{3}/\lambda_{0}$,
i.e. it depends on how the ratio between the quartic terms compares
with the ratio between the cubic terms. Importantly, both ratios may
be comparable even if $\lambda_{0},\lambda_{3}\ll u,g$. This analysis
reveals that the role of the electromagnetic fluctuations on multi-component
superconductors is potentially much more significant than in the case
of single-component superconductors.

It is important to discuss the limitations of our approach. In order
to integrate out the electromagnetic field fluctuations, we assumed
that, in the temperature range where these fluctuations are significant,
the spatial variation of the superconducting order parameter can be
neglected. Formally, this can only be justified in type-I superconductors,
for which the correlation length is smaller than the coherence length \cite{HalperinLubenskyMa1974}.
Other methods that do not require this approximation of a uniform
superconducting order parameter were also employed for the cases of
the $s$-wave and isotropic multi-component superconductors to study
the stability of the predicted first-order transition.
Perturbative $4-\varepsilon$ renormalization-group calculations and
large-$N$ expansions found the same first-order transition as in
the approach where the gauge-field fluctuations are integrated out \cite{HalperinLubenskyMa1974,Millev1990,LiBelitzToner2009}.
However, Monte Carlo simulations and duality mappings revealed a second-order
transition for type-II superconductors \cite{DasguptaHalperin1981,Kleinert1982,Sudbo2002}, indicating that a tricritical
point should take place as the ratio between the penetration depth
and coherence length is continuously changed. This was also seen in the
$d=3$ renormalization-group calculations of Refs. \cite{Herbut1996,Kleinert2003}. The implications
of these other results to our findings deserve further investigation.
As discussed above, the central point in our paper is not the first-order
nature of the transition in two-component superconductors, but the
fact that the gauge-field fluctuations affect differently the nematic
and the chiral states. Since this result is rooted on the anisotropy
of the superconducting stiffness, it is reasonable to expect that
it will play a role in the selection of the leading instability regardless
of the ratio between the penetration depth and the coherence length.
This expectation can be verified directly by appropriate Monte Carlo
simulations \cite{Sudbo2002}.

Notwithstanding these caveats, it is useful to discuss possible nematic
superconductors for which our results may be relevant. In the case
of the tetragonal iron-based superconductors Ba$_{1-x}$K$_{x}$Fe$_{2}$As$_{2}$ \cite{LiJun2017Nssi} and LiFeAs \cite{Borisenko2020}, which have
been proposed to display a spontaneous nematic superconducting state,
the scenario put forward involves nearly-degenerate $s$-wave and
$d$-wave states, for which our analysis is not applicable. Similarly,
for few-layer NbSe$_{2}$, the twofold anisotropy observed experimentally
in the superconducting state has been associated with a strain and
magnetic-field promoted admixture between $s$-wave and $d$-wave/$p$-wave
states \cite{hamill2020unexpected,cho2020distinct}, although a spontaneous condensation of a two-component
superconducting order parameter cannot be completely ruled out \cite{Shaffer2020}.
On the other hand, doped Bi$_{2}$Se$_{3}$ \cite{MatanoK2016Ssbi,ShingoYonezawa2016Tefn,PanY.Rsbi,AsabaTomoya2017RSBi},
which has a trigonal crystal structure, has been proposed to be a
nematic two-component superconductor. In this case, based on our results
from Sec. \ref{s:triangular}, gauge-field fluctuations could provide
a mechanism to stabilize a nematic superconducting state -- in addition
to the previously discussed mechanism enabled by the spin-orbit coupling \cite{Fu2014}.
As for CaSn$_{3}$, little is known about the mechanism behind the
possible nematic superconducting state reported in Ref. \cite{Siddiquee2022}. Although
its crystal structure is cubic, which was not explicitly analyzed
in this paper, we expect that the same effects uncovered for the triangular
and tetragonal lattices should emerge in this case as well. 

Finally, twisted bilayer graphene was also recently shown to
display a nematic superconducing state \cite{cao2020nematicity} (for an alternative perspective, see Ref. \cite{Sentef2021}). One proposed scenario is that
it arises from nearly-degenerate superconducting states which, in
turn, are expected from pairing either promoted by interactions involving
the van Hove points \citep{ChichinadzeDmitryV2020Nsit} or mediated
by the exchange of SU(4) spin-valley fluctuations \citep{WangYuxuan2020Tans}.
Below the degeneracy point, e.g. between $i$-wave and $d$-wave or
between $p$-wave and $f$-wave instabilities, the coexistence state
spontaneously breaks threefold rotational symmetry under certain conditions
on the system parameters (see also \cite{SZLin2018,Scheurer2020}). Alternatively, a two-component superconductor yielding a nematic superconducting
state has also been proposed \cite{FernandesVenderbos2018,lake2022pairing}. In this context, it has been shown that coupling to strong
normal-state density-wave fluctuations can promote the nematic over
the chiral state \cite{KoziiVladyslav2019Nssb}. While the effect of gauge-field fluctuations may
be relevant, a direct application of our results to twisted bilayer graphene is complicated
by the fact that this is a 2D superconductor with rather unique properties.
Indeed, as discussed in Refs. \cite{cao2020nematicity,park2021magic,MacDonald2021}, unlike most 2D superconductors, orbital
effects are significant even when in-plane magnetic fields are applied,
as the Fermi surfaces associated with opposite valleys are strongly
distorted by the in-plane fields due to inter-layer electronic tunneling.
Interestingly, in twisted multi-layer graphene with alternating twist
angles, this orbital effect is suppressed and the nematic superconducting
state is replaced by an isotropic state \cite{park2021magic}. While it is tempting to speculate
that this behavior may be attributed to a transition from nematic
to chiral superconductivity as the number of layers increases, which
should affect the impact of the gauge-field fluctuations, further
investigations are needed both theoretically and experimentally.
\begin{acknowledgments}
We thank C. Batista, A. Chubukov, and J. Schmalian for fruitful discussions.
This work was supported by the U. S. Department of Energy, Office
of Science, Basic Energy Sciences, Materials Sciences and Engineering
Division, under Award No. DE-SC0020045.
\end{acknowledgments}

\appendix

\section{Series expansion of $\mathcal{F}_{\mathrm{EM}}[\psi]$}

\label{app:taylor} We start by repeating the expression in Eq. \eqref{eq:EMfluctuationsTerm}
for $\mathcal{F}_{\mathrm{EM}}[\psi]$:
\begin{equation}
\begin{aligned} & \mathcal{F}_{\mathrm{EM}}[\psi]=\frac{4T\Lambda^{3}}{3(2\pi)^{2}}\ln\left(\psi_{s}\right)\\
 & +\frac{T\psi_{s}^{3}}{2(2\pi)^{3}}\int_{0}^{2\pi}d\phi\int_{-1}^{1}dx\int_{0}^{\tfrac{\Lambda}{\psi_{s}}}dq\,q^{2}\ln\left(c+bq^{2}+q^{4}\right)\mbox{.}
\end{aligned}
\label{eq:original}
\end{equation}

We re-write the argument of the logarithm in terms of $a_{+}$
and $a_{-}$ 
given by Eq. \eqref{eq:a}. We have: 
\begin{equation}
\ln\left(c+bq^{2}+q^{4}\right)=\ln\left(q^{2}+a_{+}^{2}\right)+\ln\left(q^{2}+a_{-}^{2}\right)
\end{equation}

Moreover, since 
\begin{equation}
\begin{aligned} & \lim\limits _{q\to0}q^{n}\ln(q)=0\mbox{ and }\\
 & \lim\limits _{q\to0}q^{n}\ln\left(\frac{t}{q^{2}}+1\right)=0
\end{aligned}
\end{equation}
for $t\in\mathbb{C}$ and $n\in\mathbb{N}$, we can further rewrite
the integrand as 
\begin{equation}
\ln\left(c+bq^{2}+q^{4}\right)=4\ln\left(q\right)+\ln\left(\frac{a_{+}^{2}}{q^{2}}+1\right)+\ln\left(\frac{a_{-}^{2}}{q^{2}}+1\right).
\end{equation}

Therefore, the original integral of Eq. \eqref{eq:original} becomes
the sum of three terms $I_{1}$, $I_{2}$ and $I_{3}$ given by 
\begin{equation}
\begin{aligned}I_{1}= & \frac{4T\Lambda^{3}}{3(2\pi)^{2}}\ln\left(\psi_{s}\right)+\frac{4T\psi_{s}^{3}}{(2\pi)^{2}}\int_{0}^{\tfrac{\Lambda}{\psi_{s}}}dq\,q^{2}\ln\left(q\right)\mbox{,}\\
I_{2}= & \frac{T\psi_{s}^{3}}{2(2\pi)^{3}}\int_{0}^{2\pi}d\phi\int_{-1}^{1}dx\int_{0}^{\tfrac{\Lambda}{\psi_{s}}}dq\,q^{2}\ln\left(\frac{a_{+}^{2}}{q^{2}}+1\right)\mbox{ and }\\
I_{3}= & \frac{T\psi_{s}^{3}}{2(2\pi)^{3}}\int_{0}^{2\pi}d\phi\int_{-1}^{1}dx\int_{0}^{\tfrac{\Lambda}{\psi_{s}}}dq\,q^{2}\ln\left(\frac{a_{-}^{2}}{q^{2}}+1\right)\mbox{.}
\end{aligned}
\end{equation}
The integral in the first term $I_{1}$ can be evaluated in a straightforward
way; we obtain:
\begin{equation}
\begin{aligned}I_{1}= & \frac{4T\Lambda^{3}}{9(2\pi)^{2}}\left[3\ln\left(\Lambda\right)-1\right]\mbox{.}\end{aligned}
\end{equation}
Therefore, the term $I_{1}$ does not depend on the order parameter,
and as such can be neglected. As for the second and third terms, $I_{2}$
and $I_{3}$, we first focus on the integral 
\begin{equation}
\begin{aligned}J\equiv & \int_{0}^{\tfrac{\Lambda}{\psi_{s}}}dq\,q^{2}\ln\left(\frac{a^{2}}{q^{2}}+1\right)\mbox{,}\end{aligned}
\end{equation}
where $a$ could be either $a_{+}$ or $a_{-}$. We split $J$ into
three parts:
\begin{equation}
\begin{aligned}J= & \int_{0}^{a}dq\,q^{2}\ln\left(\frac{a^{2}}{q^{2}}\right)+\int_{0}^{a}dq\,q^{2}\ln\left(\frac{q^{2}}{a^{2}}+1\right)\\
 & +\int_{a}^{\tfrac{\Lambda}{\psi_{s}}}dq\,q^{2}\ln\left(\frac{a^{2}}{q^{2}}+1\right).
\end{aligned}
\end{equation}
The first term in $J$ gives 
\begin{equation}
\begin{aligned}\int_{0}^{a}dq\,q^{2}\ln\left(\frac{a^{2}}{q^{2}}\right)=\frac{2a^{3}}{9}\end{aligned}
\end{equation}
whereas the second and third terms can be expressed as an infinite
series using the logarithm Taylor expansion:
\begin{equation}
\begin{aligned}J= & \frac{2a^{3}}{9}+\int_{0}^{a}dq\,q^{2}\sum_{n=1}^{\infty}\frac{(-1)^{n-1}}{n}\left(\frac{q}{a}\right)^{2n}\\
 & +\int_{a}^{\tfrac{\Lambda}{\psi_{s}}}dq\,q^{2}\sum_{n=1}^{\infty}\frac{(-1)^{n-1}}{n}\left(\frac{a}{q}\right)^{2n}\mbox{.}
\end{aligned}
\end{equation}
Performing the integrals order by order, we find 
\begin{equation}
\begin{aligned}J
= & \left[\frac{2}{9}\!+\!\sum_{n=1}^{\infty}\frac{4(-1)^{n-1}}{4n^{2}-9}\right]\!a^{3}\!+\!\sum_{n=1}^{\infty}\frac{(-1)^{n-1}a^{2n}}{n(-2n+3)}\left(\tfrac{\psi_{s}}{\Lambda}\right)^{2n-3}\!.\end{aligned}
\end{equation}
Using the result 
\begin{equation}
\begin{aligned}\frac{2}{9}+\sum_{n=1}^{\infty}\frac{4(-1)^{n-1}}{4n^{2}-9}=-\frac{\pi}{3}\mbox{,}\end{aligned}
\end{equation}
the expression for $J$ can be further simplified to 
\begin{equation}
\begin{aligned}J= & -\frac{\pi}{3}a^{3}+\sum_{n=1}^{\infty}\frac{(-1)^{n-1}a^{2n}}{n(-2n+3)}\left(\tfrac{\psi_{s}}{\Lambda}\right)^{2n-3}.\end{aligned}
\end{equation}
Substituting this expression for $J$ in the definitions of $I_{2}$
and $I_{3}$, we obtain 
\begin{equation}
\begin{aligned} & \quad I_{2}+I_{3}= -\frac{T\psi_{s}^{3}}{48\pi^{2}}\int_{0}^{2\pi}\!d\phi\int_{-1}^{1}dx\left(a_{+}^{3}+a_{-}^{3}\right)\\
 & +\sum_{n=1}^{\infty}\frac{T\psi_{s}^{2n}}{2(2\pi)^{3}}\!\int_{0}^{2\pi}\!d\phi\int_{-1}^{1}dx\frac{(-1)^{n-1}\Lambda^{-2n+3}}{n(-2n+3)}\left(a_{+}^{2n}+a_{-}^{2n}\right)\mbox{,}
\end{aligned}
\end{equation}
which gives Eq. \eqref{eq:Fem_series} in the main text.

\bibliographystyle{apsrev4-2}
\bibliography{bibliography}

\end{document}